\documentclass[11pt,letterpaper]{article}
\pdfoutput=1
\usepackage{jheppub}
\pdfoutput=1
\usepackage{epsfig}
\usepackage{amsfonts}
\usepackage{amsmath}
\usepackage{amssymb}
\usepackage{slashed}
\usepackage{color,hyperref}

\title{Complementarity of LHC and EDMs for Exploring Higgs CP Violation}
\author[a]{Chien-Yi Chen,}
\author[a]{S.  Dawson\,}
\author[b]{and Yue Zhang\,}
\affiliation[a]{Department of Physics, Brookhaven National Laboratory,\\
Upton, New York, 11973}
\affiliation[b]{Walter Burke Institute for Theoretical Physics,
California Institute of Technology,\\ Pasadena, CA 91125}
\emailAdd{cychen@bnl.gov}
\emailAdd{dawson@bnl.gov}
\emailAdd{yuezhang@caltech.edu}
\abstract{We analyze the constraints on a CP-violating, flavor conserving, two Higgs doublet model from the measurements
of Higgs properties and from the search for heavy Higgs bosons at LHC, and show that the stronger limits typically come from the heavy
Higgs search channels.  The limits on CP violation arising from the Higgs sector measurements are complementary to those from EDM measurements.
Combining all current constraints from low energy to colliders, we set generic upper bounds on the CP violating angle which parametrizes the CP odd component in the 126 GeV Higgs boson.}

\preprint{CALT-TH-2015-009}

\begin{document}
\maketitle
\section{Introduction}
Now that the $126$ GeV Higgs boson has been discovered~\cite{Aad:2012tfa,Chatrchyan:2012ufa}, the exploration of its properties is the focus of LHC phenomenology.
The current measurements  of Higgs production and decay rates are consistent with the Standard Model (SM) predictions at the $\sim10-20\%$ level, leaving open the 
possibility that there is additional physics in the Higgs sector.  One attractive alternative to the SM is the two Higgs doublet model (2HDM), 
which has $5$ Higgs bosons, allowing for new phenomena in the Higgs sector~\cite{Branco:2011iw}.  The couplings of the Higgs bosons
to fermions and gauge bosons in the 
CP conserving 2HDM depend on $2$ parameters: $\alpha$, which describes the mixing in the neutral Higgs boson sector, and $\tan\beta$, the ratio
of Higgs vacuum expectation values.  Measurements of Higgs coupling properties 
in the CP conserving limit 
require that the 2HDM be close to the alignment
limit, $\beta-\alpha\sim {\pi\over 2}$~\cite{newatlas,Chen:2013kt,Coleppa:2013dya, Cheung:2013rva}.

The SM explains CP violation through the CKM mixing matrix, which is sufficient to account for observed CP violation in the $B$ and $K$ systems.  However,
it is insufficient to explain the excess of matter over anti-matter in the universe, suggesting that there may be further sources of CP 
violation~\cite{Shu:2013uua,Morrissey:2012db}. 
The 2HDM offers the possibility for a new source of CP violation beyond the CKM matrix and QCD $\theta$ term.  
In such a scenario, the $126$ GeV Higgs boson can be a mixture of CP even and CP 
odd states~\cite{Inoue:2014nva,Brod:2013cka,Fontes:2015mea,Barroso:2012wz,Lavoura:1994fv} .  
The LHC data has already excluded the case that the 126 Higgs is a pure CP odd scalar~\cite{CMSspinparity, ATLASspinparity}, but the constraints on
its CP odd mixture are still rather weak.
There have been proposals of new techniques to directly measure the Higgs CP mixture in future colliders~\cite{Harnik:2013aja, Bishara:2013vya, Chen:2014ona, Dolan:2014upa, Berge:2014sra, He:2014xla, BarShalom:1995jb, Atwood:2000tu}.
The parameters of the CP violating version of the 2HDM receive complementary
limits from LHC Higgs coupling measurements and from low energy measurements such as electric dipole moments (EDMs).  
 The measurements of Higgs couplings do not put a strong 
constraint on the CP violating phase, especially in the alignment limit~\cite{Inoue:2014nva}, and the strongest limits come from 
EDMs~\cite{Shu:2013uua, Inoue:2014nva,Bian:2014zka,Brod:2013cka,Ipek:2013iba,Cheung:2014oaa}.

CP violation in the Higgs sector has
 been studied extensively in the MSSM limit of the 2HDM~\cite{Carena:2002bb,Bechtle:2013wla}.  The MSSM contains many sources of CP
violation from the soft SUSY breaking terms in the effective Lagrangian~\cite{Arbey:2014msa}.  The primary restriction on this type of CP
violation arises from the requirement that the lightest Higgs boson have a mass near $126$ GeV~\cite{Li:2015yla}.   Analogous limits
to those obtained in this work from Higgs couplings, heavy Higgs searches, and EDMs can be found  in the MSSM~\cite{Chakraborty:2013si,Carena:2014nza}.

We consider a CP violating 2HDM scenario which has a softly broken $Z_2$ symmetry which avoids large flavor changing neutral currents from
Higgs exchange, but allows for new CP violation from the scalar potential.    We further allow the Higgs couplings to have small deviations from the alignment limit. 
 In this work, we consider the additional constraints  on the parameters of the theory arising from the search for heavy Higgs bosons.
  In the CP conserving 2HDM, the search for heavy
 Higgs bosons significantly restricts the allowed parameter space for small $\tan\beta$ \cite{Chen:2013rba,Craig:2013hca}
 and this remains true in the CP violating case.  
In the context of the 2HDMs,  if there is significant CP violation, the heavy Higgs  boson masses cannot be too 
heavy  and in some regions of parameter space the LHC heavy Higgs searches can place the
leading constraint on CP violation.  

In Section~\ref{review}, we review the CP violating 2HDM and predictions for Higgs boson production and decay within this class of models.  Limits from heavy
Higgs searches are discussed in Section~\ref{plotsec} and compared with low energy limits from the electron EDM.  We have also updated the results
of Refs.~\cite{Freitas:2012kw, Djouadi:2013qya, Chang:2013cia, Inoue:2014nva} for the limits on the CP violating parameters from Higgs coupling fits.  
Finally, Section~\ref{sec:conc} contains a concluding discussion of the complementary limits on CP violating 2HDMs from Higgs coupling fits, heavy Higgs searches, EDMs, the oblique parameters, and $g-2$.

\label{introsec}

\section{Two Higgs Doublet Models and CP Violation}
\label{review}
In this section we review the 2HDMs considered in this study. 
\subsection{Scalar Potential with Two Higgs Doublets}
The most general two Higgs doublet potential  which breaks $SU(2)_L\times U(1)$ to $U(1)_{EM}$ is,
\begin{eqnarray}                           \label{pot}
V(\phi_1, \phi_2)&=&-\frac{1}{2}\left[m_{11}^2(\phi_1^\dagger\phi_1)
+\left(m_{12}^2 (\phi_1^\dagger\phi_2)+{\rm h.c.}\right)
+m_{22}^2(\phi_2^\dagger\phi_2)\right] 
\nonumber \\
&&+ \frac{\lambda_1}{2}(\phi_1^\dagger\phi_1)^2
+\frac{\lambda_2}{2}(\phi_2^\dagger\phi_2)^2+\lambda_3(\phi_1^\dagger\phi_1) (\phi_2^\dagger\phi_2) 
+\lambda_4(\phi_1^\dagger\phi_2) (\phi_2^\dagger\phi_1) 
\nonumber \\
&&+\frac{1}{2}\left[\lambda_5(\phi_1^\dagger\phi_2)^2 + \lambda_6 (\phi_1^\dagger\phi_2) (\phi_1^\dagger\phi_1) 
+ \lambda_7 (\phi_1^\dagger\phi_2) (\phi_2^\dagger\phi_2) +{\rm h.c.}\right] \ . 
\label{pot_gen}
\end{eqnarray}
The potential of Eq.~(\ref{pot_gen}) leads to tree level flavor changing neutral currents, which can be avoided by imposing a $Z_2$ symmetry under which,
\begin{equation}
\phi_1\rightarrow -\phi_1\qquad \phi_2\rightarrow \phi_2\, .
\label{z2sym}
\end{equation}
Eq.~(\ref{z2sym}) implies $\lambda_6=\lambda_7=0$, while a non-zero $m_{12}$ softly breaks the  $Z_2$ symmetry of Eq.~(\ref{z2sym}).  

After electroweak symmetry breaking, the Higgs doublets in 
unitary gauge can be written as,
\begin{eqnarray}\label{VEVs}
\phi_1=\begin{pmatrix}
-\sin\beta H^+ \\
\frac{1}{\sqrt2} (v \cos\beta + H_1^0 - i \sin\beta A^0)
\end{pmatrix}, \ \ 
\phi_2=e^{i\xi} \begin{pmatrix} 
\cos\beta H^+ \\
\frac{1}{\sqrt2} (v \sin\beta + H_2^0 + i \cos\beta A^0) 
\end{pmatrix} \, ,
\end{eqnarray}
where $\tan\beta=v_2/v_1$, $v=\sqrt{\mid v_1\mid^2+\mid v_2\mid^2}=246$ GeV and $H^+$ is the physical charged Higgs with mass $m_{H^+}$.  
We are free to redefine fields and go to a basis where $\xi=0$.
In general there are $2$ independent phases and 
 the imaginary parts of $m_{12}$ and $\lambda_5$ 
  lead to mixing in the neutral Higgs sector between $H_{1}^0, H_2^0$ and $A^0$, and that is the source of CP violation.
  
The mixing among  the three neutral scalars can be parametrized by an orthogonal matrix $R$, 
\begin{eqnarray}\label{R}
R =\begin{pmatrix}
-s_{\alpha}c_{\alpha_b} & c_{\alpha}c_{\alpha_b} & s_{\alpha_b} \\
s_{\alpha}s_{\alpha_b}s_{\alpha_c} - c_{\alpha}c_{\alpha_c} & -s_{\alpha}c_{\alpha_c} - c_{\alpha}s_{\alpha_b}s_{\alpha_c} & c_{\alpha_b}s_{\alpha_c} \\
s_{\alpha}s_{\alpha_b}c_{\alpha_c} + c_{\alpha}s_{\alpha_c} & s_{\alpha}s_{\alpha_c} - c_{\alpha}s_{\alpha_b}c_{\alpha_c} & c_{\alpha_b}c_{\alpha_c}
\end{pmatrix} \ .
\end{eqnarray}
where $s_\alpha=\sin\alpha$, etc and
\begin{equation}
-{\pi\over 2}< \alpha_b\le{\pi\over 2}\qquad -{\pi\over 2}\le\alpha_c\le{\pi\over 2}\, .
\end{equation}
The physical mass eigenstates are then defined as $(h_1, h_2, h_3)^T = R (H_1^0, H_2^0, A^0)^T$. 
In the CP conserving version of the 2HDM, $\alpha_b=\alpha_c=0$, $R$
is block diagonal, and $h_1$ and $h_2$ have no pseudoscalar component. 

\subsection{Neutral Scalar Interactions}
For simplicity, we focus on the 2HDMs where the Yukawa sector 
has a $Z_2$ symmetry and  $\phi_1$ and $\phi_2$ each only gives mass to up or down type fermions. This 
is sufficient to suppress tree-level flavor changing processes mediated by the neutral Higgs scalars.
For the $3rd$ generation (and suppressing CKM mixing),
\begin{eqnarray}                           
    \label{Yuk}
\mathcal{L}= \left\{\begin{array}{ll} 
-\biggl( \displaystyle{\cos\alpha\over \sin\beta}{m_t\over v} \biggr)\overline Q_L (i\tau_2) \phi_2^* t_R 
-\biggl(  {\cos\alpha\over \sin\beta}{m_b\over v}\biggr) \overline Q_L \phi_2 b_R + {\rm h.c.} & \hspace{1cm} {\rm Type\ I}\vspace{0.2cm} \\
-\biggl(  \displaystyle{\cos\alpha\over \sin\beta}{m_t\over v} \biggr)\overline Q_L (i\tau_2) \phi_2^* t_R 
+\biggl( {\sin\alpha\over \cos\beta}{m_b\over v} \biggr)\overline Q_L \phi_1 b_R
+ {\rm h.c.} & \hspace{1cm} {\rm Type\ II} \, ,
\end{array} \right.
\end{eqnarray}
where $Q_L^T=(t_L,b_L)$.
In both cases, we assume that the charged lepton Yukawa coupling has the same form as that of the charge $-1/3$  quarks.
Under the $Z_2$ symmetry, $Q_L, t_R, \phi_2$ are always even, $\phi_1$ is always odd, and $b_R$ is even (odd) in Type I (II) models.

From this we can derive the couplings between neutral Higgs bosons and the fermions and gauge bosons. As a general parametrization
we take,
\begin{eqnarray}
\mathcal{L} = \sum_{i=1}^3 \left[-m_f\left( c_{f,i} \bar f f+ \tilde c_{f,i} \bar f i\gamma_5 f  \right) \frac{h_i}{v} + \left( 2 a_i M_W^2 W_\mu W^\mu + a_i M_Z^2 Z_\mu Z^\mu \right) \frac{h_i}{v}\right] \ .
\label{coup_f}
\end{eqnarray}
When $c_{f,i}\tilde c_{f,i}\neq0$ or $a_{i}\tilde c_{f,i}\neq0$, the mass eigenstate $h_i$ couples to both CP even and CP odd operators, so the theory violates CP.
The coefficients $c_{f,i}$, $\tilde c_{f,i}$ and $a_i$ can be derived from $\tan\beta$ and the elements of the matrix $R$ defined above.
%\begin{eqnarray}\label{Hcouplings}
\begin{table}[h]
\centering{\begin{tabular}{|c|c|c|c|c|c|}
\hline
&  $c_{t,i}$ & $c_{b,i}=c_{\tau,i}$ & $\tilde c_{t,i}$ & $\tilde c_{b,i}=\tilde c_{\tau,i}$ & $a_i$ \\
\hline
Type I & $R_{i2}/\sin\beta$ & $R_{i2}/\sin\beta$ & $-R_{i3}\cot\beta$ & $R_{i3}\cot\beta$ & 
$R_{i2}\sin\beta+R_{i1}\cos\beta$ \\
\hline
Type II & $R_{i2}/\sin\beta$ & $R_{i1}/\cos\beta$ & $-R_{i3}\cot\beta$ & $-R_{i3}\tan\beta$ & 
$R_{i2}\sin\beta+R_{i1}\cos\beta$ \\
\hline
\end{tabular}} \ \ \ 
\caption{Fermion and gauge boson couplings to Higgs mass eigenstates.}
\label{Hcouplings}
\end{table}
%\end{eqnarray}
An appealing feature is that all couplings in Table \ref{Hcouplings} depend on only four parameters, $\alpha$, $\alpha_b$, $\alpha_c$ and $\tan\beta$.  
It is worth noting that the couplings of the light Higgs boson $h_1$ to the gauge bosons and fermions are independent of $\alpha_c$. 
Fits to the CP conserving 2HDM suggest that the couplings are close to the alignment limit, $\beta-\alpha\sim{\pi\over 2}$, implying that $h_1$ has  couplings 
very close to the SM predictions.  In our numerical studies, we will allow small deviations from the alignment limit. 

\subsection{CP Violation Implies a Non-Decoupled Heavy Higgs Sector}\label{sec:2.3}

In general, the imposed $Z_2$ symmetry in the Yukawa sector is not preserved by renormalization.
The hard breaking $\lambda_6, \lambda_7$ terms from the Higgs potential will induce couplings of $\phi_1$, $\phi_2$ to both up and down type quarks. This does not reintroduce any tree level flavor changing effects because the induced Yukawa matrices are still aligned with the corresponding fermion mass matrices.
%although it upsets the assumed the Yukawa structure.
Motivated by this, a convenient choice is to forbid the $\lambda_6, \lambda_7$ terms. 
In this case, the model has an approximate $Z_2$ symmetry, which is only softly broken by the $m_{12}^2$ term.

For the approximate $Z_2$ symmetric model, all of  the potential parameters can be solved for in
terms of   the following parameters:
\begin{itemize}
\item The scalar masses, $m_{h_1}$, $m_{h_2}$, $m_{h_3}$ and $m_{H^\pm}$
\item The neutral scalar mixing angles, $\alpha$, $\alpha_b$, $\alpha_c$
\item The ratio of vev's, $\tan\beta$
\item One potential parameter, ${\rm Re}(m_{12}^2)$, or $\nu \equiv {\rm Re}(m_{12})^2/({{v^2\sin2\beta}})$\, , 
\end{itemize}
giving $9$ physical parameters.
The $\nu$ parameter controls the decoupling limit, {\it i.e.}, when ${\rm Re}(m_{12})^2$ approaches  infinity, 
the masses of $h_2$, $h_3$ and $H^\pm$ also go to infinity.

The explicit solution for the parameters of the scalar potential was found in Ref.~\cite{Inoue:2014nva}, and is  listed below in Appendix~\ref{appa}. 
The imaginary part of $\lambda_5$, which is a source of CP violation, is given by,
\begin{eqnarray}
{\rm Im}\lambda_5 &=& \frac{2 \cos\alpha_b}{v^2 \sin\beta} 
\left[ (m_{h_2}^2-m_{h_3}^2) \cos\alpha \sin\alpha_c \cos\alpha_c   \rule{0mm}{4.5mm}\right. \nonumber \\
&&\hspace{3.2cm}
\left. +(m_{h_1}^2 - m_{h_2}^2 \sin^2\alpha_c-m_{h_3}^2\cos^2\alpha_c) \sin\alpha \sin\alpha_b \rule{0mm}{4.5mm}\right] \ .
\end{eqnarray}
An important point here is that, in order for the 126 GeV Higgs boson to have CP violating couplings, the heavy Higgs states must not decouple. Otherwise, the two Higgs doublet model will return to the SM limit. This is actually our main motivation for studying the bounds on the non-decoupled heavy Higgs.

Clearly, when the scalars $h_{2,3}$ are much heavier than the electroweak scale,  and $m_{h_2}\simeq m_{h_3} \equiv m_{H^+}\gg m_{h_1}$, 
\begin{eqnarray}
\mid \sin 2\alpha_b\mid  \simeq \frac{\mid {\rm Im}\lambda_5\mid  v^2}{m_{H^+}^2} \left| \frac{\sin\beta}{\sin\alpha}\right| \ .
\end{eqnarray}
The unitarity bound on ${\rm Im}\lambda_5$,
${\rm Im}\lambda_5<4 \pi$, sets the largest allowed CP violating mixing angle $\alpha_b$. 
This implies that for an ${\cal {O}}(1)~\sin \alpha_b$ to be theoretically accessible, the heavy scalars $h_2$, $h_3$ and $H^\pm$ must be not far above the electroweak scale.
 In general, for nonzero $\alpha_b$, the masses of the heavy scalars should satisfy
\begin{eqnarray}
m_{H^+} \lesssim 870\,{\rm GeV}\times \sqrt{\mid{\rm Im}\lambda_5\mid /(4\pi)} \sqrt{\mid \sin\beta/(\sin\alpha\sin 2\alpha_b)\mid }\ .
\end{eqnarray}
A similar conclusion holds when one goes beyond the approximate $Z_2$ symmetry by including the $\lambda_6, \lambda_7$ terms.

\subsection{Beyond Approximate $Z_2$ Symmetry}
\label{sec:noz2}
For the approximate $Z_2$ symmetric model, there is a further theoretical constraint on the physical parameters resulting from the minimization of
the potential.  This constraint is
given in Eq.~(\ref{ab}) and can be transformed into a quadratic equation for $\tan\alpha_c$. The condition for $\alpha_c$ to have a real solution is
\begin{eqnarray}\label{prejudice}
\sin^2\alpha_b \leq \frac{(m_{h_3}^2-m_{h_2}^2)^2 \cot^2(\alpha+\beta)}{4 (m_{h_2}^2-m_{h_1}^2) (m_{h_3}^2-m_{h_1}^2)} \equiv \sin^2\alpha_b^{\rm max} \ .
\end{eqnarray}
When Eq.~(\ref{prejudice}) is satisfied, the solutions for $\alpha_c$ are,
\begin{eqnarray}          \label{ac}              
\alpha_c = \left\{\begin{array}{ll} 
\alpha_c^-, & \hspace{0.3cm} {\rm \alpha+\beta\leq0} \\
\alpha_c^+, & \hspace{0.3cm} {\rm \alpha+\beta>0} 
\end{array} \right., \hspace{0.6cm}
\tan\alpha_c^\pm\!=\!\frac{\mp| \sin\alpha_b^{\rm max}| \!\pm\! \sqrt{ \sin^2\alpha_b^{\rm max} - \sin^2\alpha_b }}{\sin\alpha_b}
\sqrt{\frac{m_{h_3}^2-m_{h_1}^2}{m_{h_2}^2-m_{h_1}^2}}\ . \hspace{-0.8cm}\nonumber \\
\end{eqnarray}

Eq.~(\ref{prejudice}) implies an additional theoretical upper bound on the CP violating angle $\alpha_b$, when the other parameters are fixed.
In practice, we sometimes find this bound can be stronger than all the experimental limits. 
However, this is only a bound because of theoretical prejudice. 
In fact, it can be removed with a minimal step beyond the approximate $Z_2$ symmetric case by introducing a $\lambda_7$ term, with $\lambda_7$ being purely imaginary. In this case, the bound Eq.~(\ref{prejudice}) no longer exists,  $\alpha_c$ becomes a free parameter, and ${\rm Im}\lambda_7$ can in turn be solved for as,
\begin{eqnarray}
{\rm Im}\lambda_7 &=& \frac{2\cos\alpha_b}{v^2 \tan^2\beta} \left[ (m_{h_3}^2-m_{h_2}^2) \sin\alpha_c \cos\alpha_c \frac{\cos(\alpha+\beta)}{\cos^2\beta} \right. \nonumber \\
&& \hspace{1.8cm} \left.+(m_{h_2}^2 \sin^2\alpha_c + m_{h_3}^2\cos^2\alpha_c - m_{h_1}^2) \sin\alpha_b\frac{\sin(\alpha+\beta)}{\cos^2\beta} \right] \ .
\end{eqnarray}
Although introducing hard $Z_2$ breaking  ($\lambda_{6,7}\ne 0$) makes the Yukawa structure in Eq.~(\ref{Yuk}) unnatural, one might argue it is accidentally the case at the electroweak scale.~\footnote{We are aware  that allowing $Z_2$ breaking terms in the Yukawa sector can introduce additional sources of CP violation. 
The price for this is introducing tree level flavor changing effects at the same time, and some flavor alignment mechanism must be resorted to~\cite{Buras:2010zm,Cline:2011mm,Jung:2013hka}.
We do not consider such a  possibility, but focus on CP violation only from the Higgs sector in this work.}
In the phenomenological study in the next section, we will give the results for both the approximate $Z_2$ case, and the minimal extension as discussed in this subsection.

\subsection{Production and Decay of the Heavy Higgs at LHC}\label{prodecay}
\subsubsection{Production}

The dominant heavy Higgs  boson
production channels relevant to this study are gluon fusion, $gg\to h_{2,3}$, vector boson fusion, $qq\to qq h_{2,3}$,
 and production in association with bottom quarks, $gg/q\bar q\to h_{2,3} b\bar b$. 
In the 2HDM we explore, the  interactions between the heavy neutral Higgs bosons
and the SM fermions and  the $W, Z$ gauge bosons are simply rescaled from those of a SM-like Higgs boson, $H_{SM}$,  by a factor given in 
Table~\ref{Hcouplings}. Therefore, it is convenient to take the SM-like Higgs cross sections, and rescale them with these factors and the appropriate form factors. The LHC production cross sections for a heavy SM-like Higgs boson  have been calculated by the
LHC Higgs Cross Section Working Group and given in~\cite{Dittmaier:2011ti,HiggsXsec}.

For the gluon fusion process, we  calculate the ratio of the
heavy Higgs boson production cross section in a CP violating 2HDM to that of a SM-like Higgs with the same mass. At one-loop,
\begin{eqnarray}
R_{gg}^i=\frac{\sigma (gg\to h_i)}{\sigma (gg\to H_{\rm SM})} = 
\frac{\left|c_{t,i} A_{1/2}^H(\tau^i_{t}) + c_{b,i} A_{1/2}^H(\tau^i_{b})\right|^2 + \left|\tilde c_{t,i} A_{1/2}^A(\tau^i_{t}) + \tilde c_{b,i} A_{1/2}^A(\tau^i_{b})\right|^2}{\left|A_{1/2}^H(\tau^i_{t}) + A_{1/2}^H(\tau^i_{b})\right|^2},
\end{eqnarray}
where $\tau^i_f=m_{h_i}^2/(4m_f^2)$ and $i=1,2,3$, $f=t,b$. The form factors $A_{1/2}^H$, $A_{1/2}^A$ are given by
\begin{eqnarray}
A_{1/2}^H(\tau) &=& 2\left(\tau +(\tau-1) f(\tau) \right) \tau^{-2} \ ,  \\
A_{1/2}^A(\tau) &=& 2 f(\tau) \tau^{-1} \ ,  \\
f(\tau) &=& \left\{ \begin{array}{ll}
{\rm arcsin}^2\left( \sqrt{\tau} \right), & \hspace{0.5cm} \tau\leq 1 \\
\frac{1}{4} \left[ \log\left( \frac{1+\sqrt{1-\tau^{-1}}}{1-\sqrt{1-\tau^{-1}}} \right) -i\pi \right]^2, & \hspace{0.5cm} \tau> 1
\end{array}\right.\, .
\end{eqnarray}
%We take the values of the cross section $\sigma (gg\to H_{\rm SM})$ from~\cite{GGHcrosssection}.

For vector boson fusion, the ratio of the heavy Higgs production cross section in a 2HDM to that of a SM-like Higgs with the same mass is simply
\begin{eqnarray}
R_{VBF}^i=
\frac{\sigma (qq\to qq h_i)}{\sigma (qq\to qq H_{\rm SM})} = (a_i)^2 .
\end{eqnarray}
%Again we take the values of the cross section $\sigma (qq\to qq H_{\rm SM})$ from~\cite{GGHcrosssection}.

For  $h_{2,3}b\bar b$ associated production, we take the NLO cross section for SM-like Higgs boson production in the $4$ flavor number scheme
 from Ref.~\cite{Heinemeyer:2013tqa,bbHcrosssection}. 
There the cross section contains two pieces, one is proportional to $y_b^2$, and the other proportional to $y_by_t$ from
  interference. We rescale these results with the heavy Higgs-fermion couplings in a 2HDM,
\begin{eqnarray}
\sigma({b\bar b\to h_{i}}) = (c_{b,i})^2 \sigma_b^H (m_{h_i}) + c_{t,i} c_{b,i} \sigma_t^H (m_{h_i})
+(\tilde c_{b,i})^2 \sigma_b^A (m_{h_i}) + \tilde c_{t,i} \tilde c_{b,i} \sigma_t^A (m_{h_i}) \, ,
\end{eqnarray}
where $\sigma_b^H$ is the cross section for $gg\rightarrow b{\overline b}h_i$ where the Higgs couples to the $b$ quarks, $\sigma_t^H$ is the
interference between diagrams contributing to $gg\rightarrow b{\overline b}h_i$ where the Higgs couples to the $b$ and the $t$ quark.  $\sigma_b^A$
and $\sigma_t^A$ are the corresponding contributions from the pseudoscalar couplings to the $b$ and $t$ quarks
 given in Eq.~(\ref{coup_f}). Results in the $5$ flavor number scheme \cite{Maltoni:2012pa} are quite similar and do not affect our conclusions.

%where $\sigma_b^H$, $\sigma_b^A$ are obtained from files named {\tt \#.grid}, $\sigma_t^H$, $\sigma_t^A$ from files named {\tt \#.top}, and {\tt \#} means the center of mass energy at LHC.

\subsubsection{Decays}

The heavy neutral scalar to electroweak gauge boson decay rates are
\begin{eqnarray}
\Gamma(h_i \to VV) = \left( a_i \right)^2 \frac{G_F m_{h_i}^3}{16\sqrt{2}\pi} \delta_V \left( 1 - \frac{4M_V^2}{m_{h_i}^2} \right)^{1/2} 
\left[ 1 - \frac{4M_V^2}{m_{h_i}^2} + \frac{3}{4} \left( \frac{4M_V^2}{m_{h_i}^2} \right)^2 \right] \ ,
\end{eqnarray}
where $V=W,Z$ and $\delta_W=2$, $\delta_Z=1$, and $i=2,3$.
We note that in the alignment limit, $\Gamma(h_{2,3} \to VV)=0$ when $\sin\alpha_b=\sin\alpha_c=0$. These channels open up with non-zero
CP violation.
The decay rates to SM fermions are
\begin{eqnarray}
\Gamma(h_i \to \bar f f) = \left[ (c_{f,i})^2 + (\tilde c_{f,i})^2 \rule{0mm}{4mm}\right] \frac{N_cG_F m_f^2 m_{h_i}}{4\sqrt{2}\pi} \left( 1 - \frac{4m_f^2}{m_{h_i}^2} \right)^{3/2} \ ,
\end{eqnarray}
where $N_c=3$ for quarks and 1 for charged leptons. 

The heavy scalars can also decay to a pair of gluons via a loop of top or bottom quarks, and the rates are
\begin{eqnarray}
\Gamma(h_i \to gg) = \frac{\alpha_s^2 G_F m_{h_i}^3}{64\sqrt{2}\pi^3} \left[ \left|c_{t,i} A_{1/2}^H(\tau^i_{t}) +c_{b,i} A_{1/2}^H(\tau^i_{b})\right|^2 + \left|\tilde c_{t,i} A_{1/2}^A(\tau^i_{t}) + \tilde c_b^i A_{1/2}^A(\tau^i_{b})\right|^2 \right] . \ \ \ 
\end{eqnarray}
Clearly  a decay rate is a CP even quantity. Thus, in all the above decay rates, the CP even coefficient $c_f^i$ and the CP odd one $\tilde c_f^i$ always contribute incoherently.

In our study, we are also interested in the heavy neutral scalars, $h_2,h_3$,
 decaying into the $Z$ boson and the 126 GeV Higgs boson,
\begin{eqnarray}\label{gAZH}
\Gamma(h_i \to Zh_1) &=& \frac{|g_{iz1}|^2}{16\pi m_{h_i}^3} \sqrt{\left( m_{h_i}^2 - (m_{h_1}+M_Z)^2 \right)\left( m_{h_i}^2 - (m_{h_1}-M_Z)^2 \right)}  \nonumber \\
&&\times \left[ - (2m_{h_i}^2+2m_{h_1}^2-M_Z^2) +\frac{1}{M_Z^2} (m_{h_i}^2-m_{h_1}^2)^2 \right]\, ,
\end{eqnarray}
where $g_{iz1}=(e/\sin2\theta_W)\left[(-\sin\beta R_{11}+\cos\beta R_{12}) R_{i3} - (-\sin\beta R_{i1}+\cos\beta R_{i2}) R_{13} \right]$.

We have also calculated the decay rate of $h_i \to 2h_1$ from the Higgs self-interactions. The decay rate is 
\begin{eqnarray}
\Gamma(h_i \to h_1 h_1) &=& \frac{g_{i11}^2 v^2}{2\pi m_{h_i}} \sqrt{1-\frac{4 m_{h_1}^2}{m_{h_i}^2}} \ ,
\label{trihiggs}
\end{eqnarray}
where $g_{i11}, (i=2,3)$ are defined in Appendix B.

To get the branching ratios, we calculate the total width of the heavy Higgs\footnote{ The rate $h_i\rightarrow \gamma \gamma$ for $i=2,3$ is
always small and can be neglected here.},
\begin{eqnarray}
\Gamma_{tot}(h_i) &=& \Gamma(h_i \to W^+W^-) + \Gamma(h_i \to ZZ) + \Gamma(h_i \to t\bar t) + \Gamma(h_i \to b\bar b) \nonumber \\
&+& \Gamma(h_i \to \tau^+ \tau^-) + \Gamma(h_i \to gg) + \Gamma(h_i \to Z h_1) + \Gamma(h_i \to h_1 h_1) \ .
\end{eqnarray}
Finally, for each channel, the ratio of signal strengths in the 2HDM to the counterpart in the SM is given by,
\begin{eqnarray}
\mu_i^{XX}=
\frac{(\sigma_{7}^i \mathcal{L}_7 + \sigma_{8}^i \mathcal{L}_8) \times {\rm Br}(h_i\rightarrow XX)}
{(\sigma_{7}^{\rm SM}\mathcal{L}_7 + \sigma_{8}^{\rm SM}\mathcal{L}_8) \times {\rm Br}^{\rm SM}(h_i\rightarrow XX)} \,  ,
\end{eqnarray}
where, for example, the production cross sections are given by 
\begin{eqnarray}
\sigma_7^i =\sigma_{gg,7} R_{gg}^i+\sigma_{VBF,7}R_{VBF}^i+\sigma_{VH,7}R_{VH}^i\, ,
\end{eqnarray}
$\sigma_{gg,7}$ is the gluon fusion cross section from Ref.~\cite{Dittmaier:2011ti,HiggsXsec} for a SM Higgs boson with a mass of $m_{h_i}$, 
and $\mathcal{L}_{7,8}$ are the luminosities used in the experimental analysis.
With this quantity, we are able to reinterpret the constraints on a heavy SM-like Higgs boson for the  heavy neutral scalars in the 2HDM.

\subsection{CP violation and Heavy Higgs Signal Rates}\label{section2.6}

At this point, it is useful to gain some intuition about the impact of CP violation on the heavy Higgs to gauge boson decay channels, 
$h_i \to  VV$ and $h_i \to Zh_1$ with $(i=2,3)$. It is convenient to redefine the Higgs doublets and go to a basis where only one doublet, called $\phi_1'$,
 gets the 246 GeV vev, while the other $\phi_2'$ has no vev~\cite{Haber:2010bw,Lavoura:1994fv}.

We start from a special point in the parameter space where the lightest Higgs, $h_1$, has exactly the same couplings does the SM Higgs boson.
This corresponds to having the mixing angles in Eq.~(\ref{R}) satisfy $\alpha_b=\alpha_c=0$, and $\beta-\alpha=\pi/2$. 
The Higgs sector preserves CP invariance at this point. 
In this case,  $h_1$ is the excitation arising from $\phi_1'$ defined above, while $h_{2,3}$ are excitations from $\phi_2'$.
As a result, the decay rates $h_i \to  VV$ and $h_i \to Zh_1$ both vanish for $i=2,3$.
It is worth noticing that this special point can be approached without going to the real decoupling limit by sending the second doublet mass to infinity.

\begin{figure}[t]
\vspace{0.5cm}
\centerline{\includegraphics[width=0.9\columnwidth]{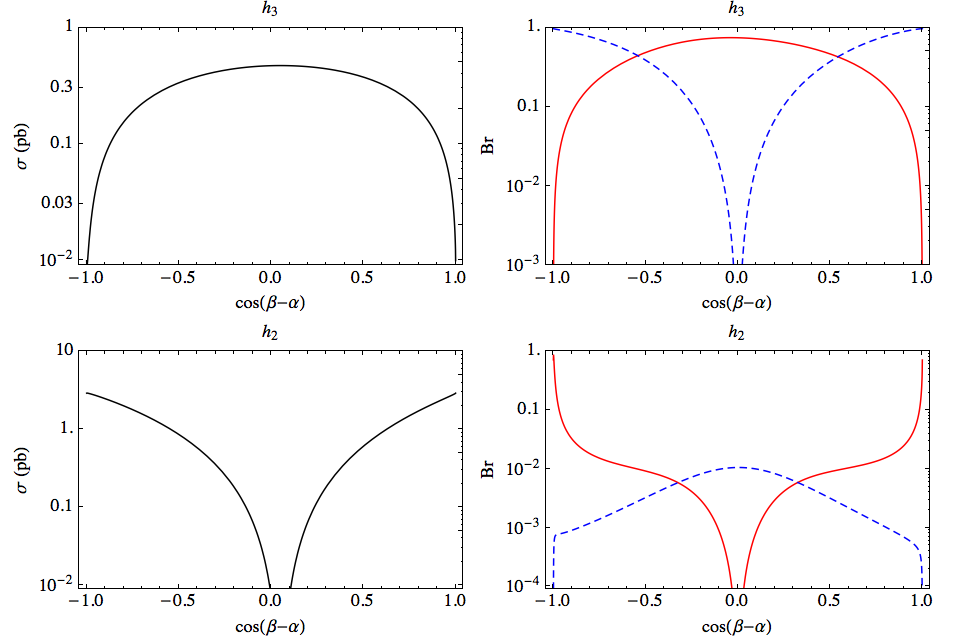}}
\caption{An example showing the impact of a non-zero
 CP violating angle, $\alpha_b=0.5$, and the deviation from
 alignment (parameterized by $\cos(\beta-\alpha)$) on the heavy Higgs production
 from gluon fusion at $\sqrt{s}=8$ TeV (left panels) and their decays (right panel) in
  $h_i \to  VV$ (red, solid) and $h_i \to Zh_1$ (blue, dashed) channels.
We have fixed the other parameters to be $\tan\beta=20$, $\alpha_c=0$, $m_{h_2}=400\,$GeV, $m_{h_3}=450\,$GeV and $\nu=1$.
} \label{sample}
\end{figure}

Next, we turn on CP violation by making $\alpha_b=0.5$, but still keep $\alpha_c=0$.
Here we discuss an example by fixing $\tan\beta=20$ (in the basis of $\{\phi_1, \phi_2\}$ given in Eq.~(\ref{VEVs})) and vary the angle $\alpha$, or the quantity $\cos(\beta-\alpha)$. 
We also choose the heavy neutral scalar masses to be $m_{h_2}=400\,$GeV and $m_{h_3}=450\,$GeV.
In Fig.~\ref{sample}, we plot the gluon fusion production cross section and the gauge boson branching ratios of $h_{2,3}$ as a function of $\cos(\beta-\alpha)$.
There are several suppressed regions which  can be understood from Table~\ref{Hcouplings}. In the case $\alpha_c=0$, we have in  the 
Type-I model,
\begin{eqnarray} \label{Csample}
&&c_{t,2} = c_{b,2} = - \frac{\sin\alpha}{\sin\beta}, \ \ \tilde c_{t,2} =-\tilde c_{b,2} = 0 \ , \nonumber \\
&&c_{t,3} = c_{b,3} = - \frac{\cos\alpha}{\sin\beta} \sin\alpha_b, \ \ \tilde c_{t,3} =-\tilde c_{b,3} = -\cos\alpha_b \cot\beta \ . 
\end{eqnarray}
First, the gluon fusion production cross section for $h_2$ via a
 top or bottom loop vanishes at $\alpha=0$ (near $\cos(\beta-\alpha)\simeq0$).
 In the example we describe here, $\beta=\arctan(20)$ is close to $\pi/2$, and  $c_{t,2} = c_{b,2}$ vanishes at $\alpha=0$.  Second, at $\alpha=\pm\pi/2$, (near $\cos(\beta-\alpha)\simeq\pm1$), the couplings $c_{t,3}$ and $c_{b,3}$ vanish. As a result, the  production cross section for $h_3$ is suppressed because $\tilde c_{t,3}$ and $\tilde c_{b,3}$ are both suppressed by $\cot\beta=1/20$ in this case.
On the other hand, the gauge boson decays of $h_{2,3}$ are directly controlled by $\beta-\alpha$. We list the relevant couplings here, again for $\alpha_c=0$,
\begin{eqnarray} \label{Asample}
&&a_2 = -\cos(\beta-\alpha), \ \ g_{2z1} = -\frac{e}{\sin2\theta_W} \sin(\beta-\alpha) \sin\alpha_b  \nonumber \\
&&a_3 = -\sin(\beta-\alpha) \sin\alpha_b, \ \ g_{3z1} = \frac{e}{\sin2\theta_W} \cos(\beta-\alpha) \ ,
\end{eqnarray}
where $g_{iz1}\ (i=2,3)$ is the coupling between $h_i$-$Z$-$h_1$ defined below Eq.~(\ref{gAZH}). 
These make it manifest why the heavy Higgs to gauge boson decay channels are sensitive both to a deviation from the alignment limit and to CP violation.
Clearly, when $\cos(\beta-\alpha)=\pm1$, the decay rates 
$h_3 \to  VV$ and $h_2 \to Zh_1$ vanish, while when $\cos(\beta-\alpha)=0$, the decay rates 
$h_2 \to  VV$ and $h_3 \to Zh_1$ vanish. 
For the case of $h_2$ decay, the branching ratios are more suppressed because the decay $h_2\to h_1 h_1$ dominates 
in most of the parameter space.
Therefore, the most important constraints come from the $h_3 \to  VV$ and $h_3 \to Zh_1$ channels.

Combining Eqs.~(\ref{Csample}) and (\ref{Asample}), we find the $h_3 \to VV$ signal rate (production cross section $\times$ decay branching ratio) is peaked at $\cos(\beta-\alpha)=0$, while 
$h_3 \to Zh_1$ vanishes at both $\cos(\beta-\alpha)=0, \pm1$, and is peaked in between.
With these facts, one can understand the yellow and orange  regions in the upper right panel of Fig.~\ref{bma1}.
One can also follow a similar analysis in order to understand the generic  features in the other plots.

\label{sec:bkgd}

\section{Results}
\label{plotsec}
In this section, we describe our method to obtain constraints from heavy Higgs searches
at the LHC, and show the numerical results in a series of figures.

In the presence of CP violation, all of  the three neutral scalars mix together, 
and we fix the lightest scalar, $h_1$, to be the 126 GeV scalar already discovered at the LHC. 
As discussed in the previous sections, 
the heavy Higgs to gauge boson decay channels, including $h_{2,3}\to WW/ZZ$ and $h_{2,3}\to Z h_1\rightarrow l^+l^- b {\overline b}$, are not only 
sensitive to deviations from the alignment limit ($\beta-\alpha=\pi/2$), but also to
the presence of CP violation ($\alpha_b, \alpha_c\neq0$).
We use the production and decay rates calculated in Sec.~\ref{prodecay} to 
obtain the 2HDM predictions for the heavy Higgs signal strength in these two channels.
Then we compare these theory predictions to the results from the 7 and 8 TeV running of
the  LHC.
For the heavy Higgs search data, we use limits for masses up to a TeV from the $h_{2,3}\rightarrow WW/ZZ$ channel~\cite{CMSH2WWZZ,TheATLAScollaboration:2013zha}  and from the $h_{2,3}\rightarrow Zh_1\rightarrow l^+l^- b {\overline b} (\tau^+\tau^-)$ channel~\cite{CMSA2Zh,Aad:2015wra}.

\begin{figure}[t]
\centerline{\includegraphics[width=0.5\columnwidth]{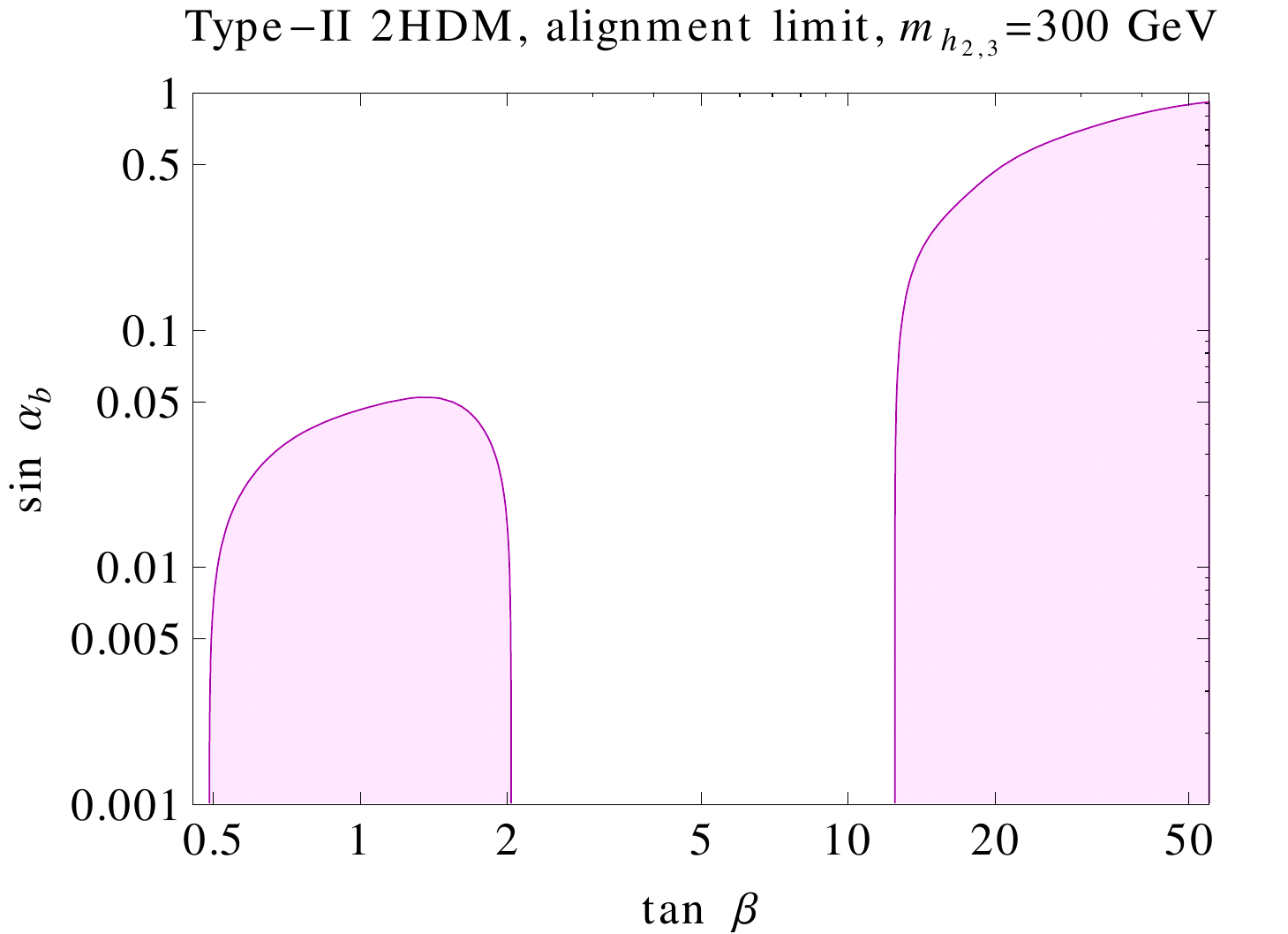}}
\caption{Heavy Higgs search constraints on the Type-II 2HDM with approximate $Z_2$ symmetry and $m_{h_2}= m_{h_3} =300$ GeV. 
The colored regions are excluded by the search for $h_{2,3}\to \tau^+\tau^-$.}
\label{fg:htautau}
\end{figure}

We also take into account the $h_{2,3} \to \tau^+\tau^-$ channel~\cite{CMSA2tautau}, which gives constraints for heavy Higgs masses up to 
a TeV and is relevant in the Type-II model in the large $\tan\beta$ case~\cite{Djouadi:2013vqa}.
 The experimental results are given as correlated fits to
  $\sigma(b{\overline b}\rightarrow h) BR(h\rightarrow \tau^+\tau^-)$ versus
 $\sigma(gg\rightarrow h)BR(h\rightarrow \tau^+\tau^-)$,  which are shown in Fig. 8 of 
  Ref.~\cite{CMSA2tautau}. 
   We interpret these bounds as bounds on the production of heavy Higgs states in the CP violating 2HDM. 
For the case $m_{h_2}= m_{h_3} =300$ GeV, the limits are shown in Fig.~\ref{fg:htautau}\footnote{We assume that 
there is no interference between the $h_2$ and $h_3$ resonances.}. In addition to a limit for $\tan\beta\gtrsim 30$, there is another limit  
around $\tan\beta \sim 1$. This lower limit arises
 because  the masses of $h_{2,3}$ are below the $t \bar{t}$ threshold and hence the dominant decay channel is through
$h_{2,3} \to b \bar{b}$ and $h_{2,3} \to \tau^+ \tau^-$. For $m_{h_2}= m_{h_3} =500$ GeV the bound from $h_{2,3} \to \tau^+\tau^-$ gives no constraint for 
$\tan\beta<50$. In general, we find that the bounds from $h_{2,3} \to \tau^+\tau^-$ are always weaker than those from the coupling measurement of the light Higgs or EDMs. As a result, we will not include them in the following plots.

The most up-to-date 126 GeV Higgs coupling data are given in Table~\ref{dat}, normalized to the appropriate luminosities. 
They are used to constrain the theoretical predictions for the  signal rates of $h_1$, from Sec.~\ref{prodecay}.
We take the SM cross sections from the LHC Higgs Cross Section Working Group~\cite{HiggsXsec}.
We have performed a $\chi^2$ analysis using the results listed in Table~\ref{dat}.
\begin{table}[h]
\caption{ATLAS and CMS Higgs Coupling Measurements.}
\centering
\begin{tabular}[t]{|l|c|c|c|c|}
\hline\hline
Channel& $\mu_{CMS}$& Ref. & $\mu_{ATLAS}$ & Ref.\\
\hline
$\mu_{WW}$& $0.83\pm 0.21$& \cite{CMS126Higgs}& $1.09^{+0.23}_{-0.21}$ &\cite{ATLAS:2014aga}\\
$\mu_{ZZ}$ & $1.0\pm 0.29$ & \cite{CMS126Higgs}& $1.44^{+0.40}_{-0.33}$ &\cite{ATLAS126HiggsZZ} \\
$\mu_{\gamma\gamma}$ & $1.13\pm0.24$ & \cite{CMS126Higgs}& $1.17\pm 0.27$ &\cite{Aad:2014eha}\\
$\mu_{bb}$ & $0.93\pm 0.49$ & \cite{CMS126Higgs}& $0.5\pm 0.4 $ &\cite{ATLAS126Higgsbb} \\
$\mu_{\tau\tau}$ & $0.91\pm 0.27$ & \cite{CMS126Higgs}& $1.4\pm 0.4$ &\cite{Aad:2015vsa}\\
\hline
\end{tabular}
\label{dat}
\end{table}
%CMS from CMS-PAS-HIG-14-09, ATLAS from web updated 12/1/15.  
%ATLAS with $4.5-4.7~ fb^{-1}$ at 7 TeV, $20.3 fb^{-1}$ at 8 TeV, with $m_h=125.36$.
%ATLAS $\gamma \gamma$ from 1408.7084 has 5.4 fb-1 at 7 TeV, 20.3 fb-1 at 8 TeV.

%\end{document}

\begin{figure}[t]
\vspace{0.5cm}
\centerline{\includegraphics[width=1.0\columnwidth]{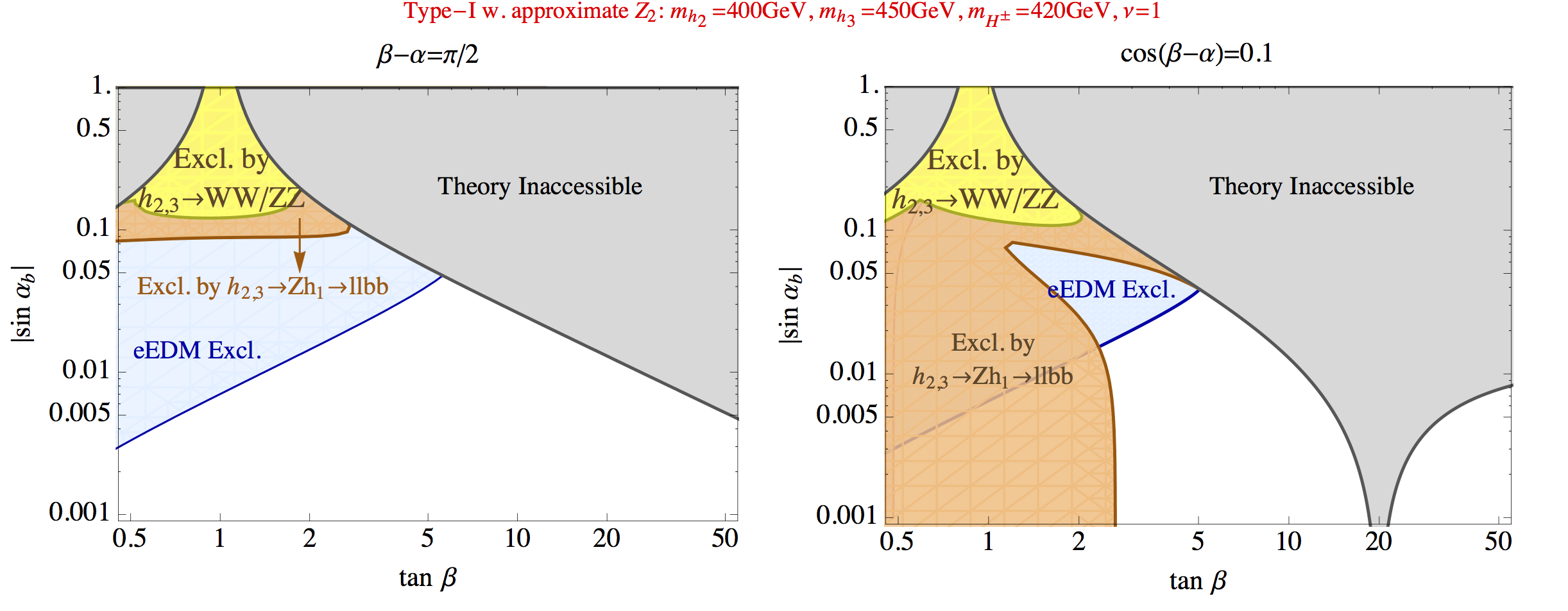}}
\caption{Heavy Higgs search constraints on the Type-I 2HDM with approximate $Z_2$ symmetry, using the $h_{2,3}\to WW/ZZ$ (yellow) and $h_{2,3}\to Z h_1\to l^+l^-b\bar b$ (orange) channels. These constraints are presented in the $\sin\alpha_b$ versus $\tan\beta$ parameter space
and colored regions are excluded. 
The left panel is for the alignment limit with $\alpha=\beta-\pi/2$, while the right panel shows the case with a deviation from that limit.
Also shown in blue are the electron EDM excluded regions.
In these plots, we have chosen the heavy scalar masses to be $m_{h_2}=400\,$GeV, $m_{h_3}=450\,$GeV, $m_{H^+}=420\,$GeV, and the model parameter $\nu=1$. The other mixing angle $\alpha_c$ is a dependent quantity fixed by Eq.~(\ref{ac}).
In the gray region, there is no real solution for $\alpha_c$.
} \label{z2t1}
\end{figure}

\begin{figure}[t]
\centerline{\includegraphics[width=1.0\columnwidth]{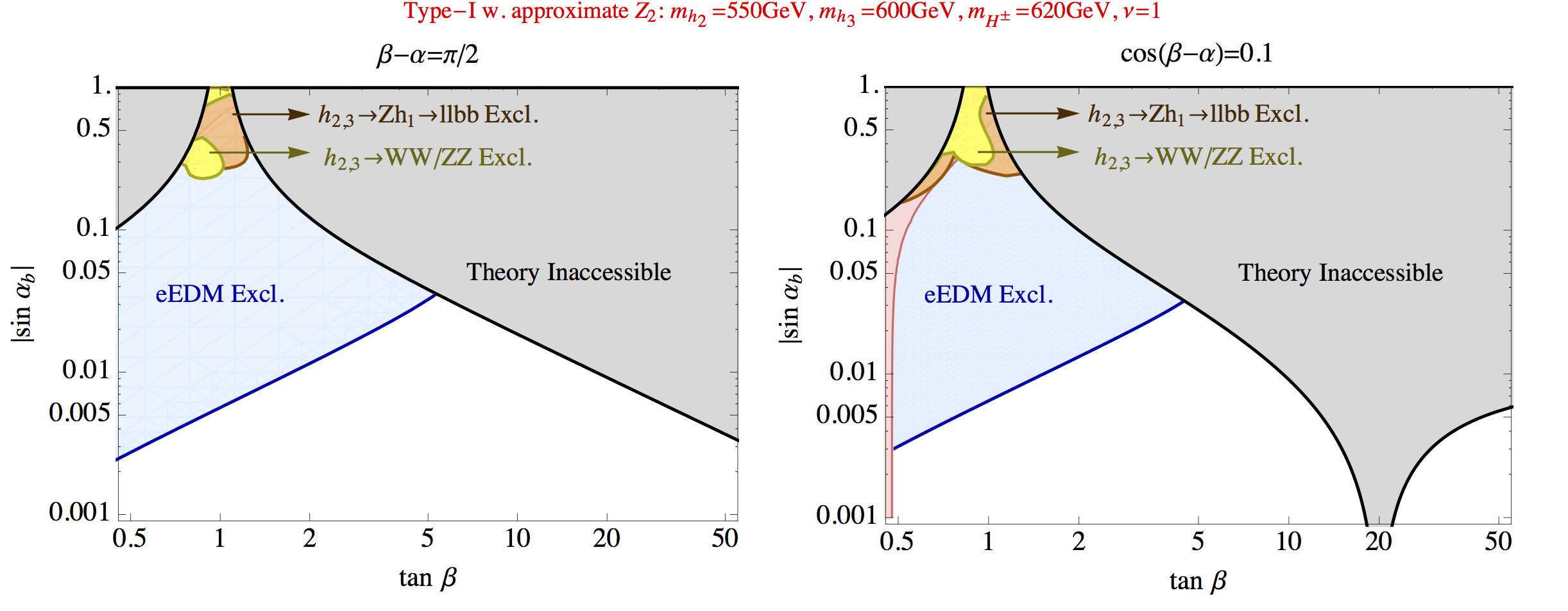}}
\caption{Similar to Fig.~\ref{z2t1}, but with heavy scalar masses $m_{h_2}=550\,$GeV, $m_{h_3}=600\,$GeV, $m_{H^+}=620\,$GeV.
In the right panel, the red region is excluded by the 126 GeV Higgs data applied to $h_1$.}
\label{fg:z2t1600}
\end{figure}

\begin{figure}[t]
\centerline{\includegraphics[width=1.0\columnwidth]{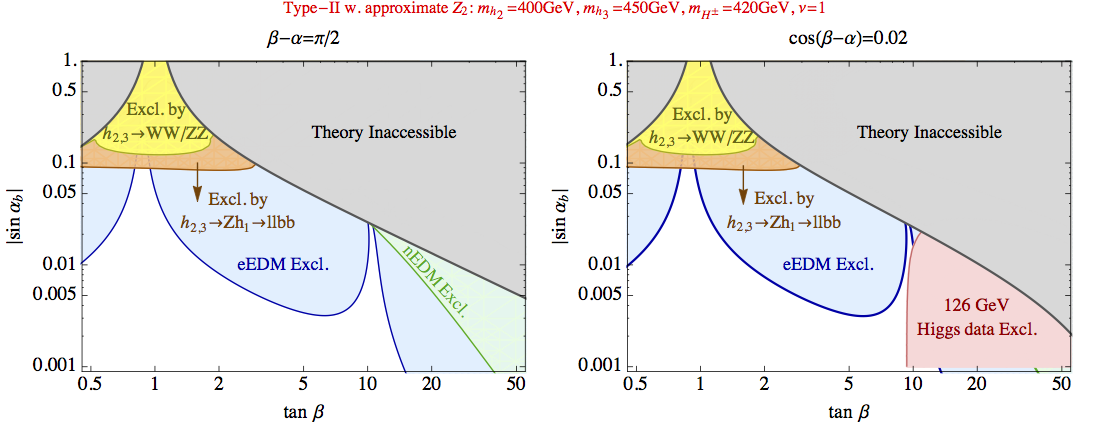}}
\caption{Heavy Higgs search constraints on the Type-II 2HDM with approximate $Z_2$ symmetry. 
The Higgs sector parameters are chosen to be the same as those in Fig.~\ref{z2t1}. The colored regions are excluded by
searches for $h_{2,3}\to WW/ZZ$ (yellow), $h_{2,3}\to Z h_1\to l^+l^-b\bar b$ (orange) channels,  
126 GeV Higgs coupling data (red), electron EDM measurements(blue), and neutron EDM limits(green). 
The gray region is again theoretically excluded because it contains no real solution for $\alpha_c$.}
\label{fg:z2t2}
\end{figure}

\begin{figure}[t]
\centerline{\includegraphics[width=1.0\columnwidth]{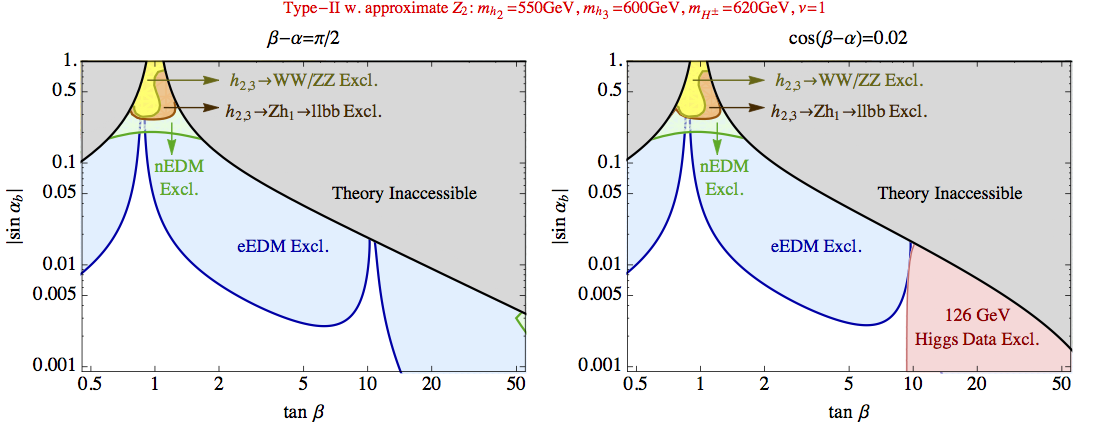}}
\caption{Similar to Fig.~\ref{fg:z2t2}, but with heavy scalar masses $m_{h_2}=550\,$GeV, $m_{h_3}=600\,$GeV, $m_{H^\pm}=620\,$GeV.}
\label{fg:z2t2600}
\end{figure}

In Figs.~\ref{z2t1} to \ref{bma2}, we show the limits derived from heavy Higgs searches and the light (126 GeV) Higgs data, 
together with those from the low energy electron and neutron EDMs. For the EDM constraints, we use the results of Ref.~\cite{Inoue:2014nva}.

In these numerical results, we fix the heavy Higgs masses and the parameter $\nu=1$. 
The CP violating angle $\alpha_c$ is fixed in the approximate $Z_2$ symmetric model by Eq.~(\ref{ac}). 
On the other hand, for the extended model without an approximate $Z_2$ symmetry, $\alpha_c$ is a free parameter.
We also note that varying the parameter $\nu$ between 0 and 1 only leads to slight changes to our results.
The constraints are shown in the $\sin\alpha_b$ versus $\tan\beta$ plane, while varying $\alpha$ and $\alpha_c$.
We consider both the alignment limit with $\alpha=\beta-\pi/2$
and cases when there are small deviations from alignment, $\cos(\beta-\alpha)=\pm \Delta$. 
The 126 GeV Higgs data put upper bounds on $\Delta$ for fixed values of $\tan\beta$.  For the Type -I model, we  consider $\Delta=0.1$, while for 
the Type- II model,  the light Higgs coupling data constraint is stronger at large $\tan\beta$, 
so we take $\Delta=0.02$~\cite{newatlas}. ATLAS has also limited the parameters of the 2HDM by directly searching for the heavier neutral Higgs boson,
but these limits are not competitive with the 
Higgs 
coupling data for the heavy $h_{2,3}$ masses that we consider~\cite{ATLAS:2013zla} .

\subsection{Limits from Heavy Higgs Searches in Approximate $Z_2$ Symmetric Models}

We first discuss the models with an approximate $Z_2$ symmetry.
Fig.~\ref{z2t1} shows the limits on the CP violating parameter, $\alpha_b$, as a function
of $\tan\beta$ in the Type-I model.  
In each panel, the gray area marked ``theory inaccessible'' has no real
solution for $\alpha_c$ from Eq.~(\ref{ac}). 
The left panel assumes the alignment limit, $\beta-\alpha\sim {\pi/2}$, while
the right panel allows for a small deviation from the alignment limit\footnote{The results are similar for negative $\Delta$.}.  The orange area
is excluded by the heavy Higgs search channel $h_{2,3}\rightarrow Zh_1\rightarrow l^+l^- b {\overline b}$,
while the yellow area is excluded by the channel $h_{2,3}\rightarrow WW/ZZ$. 
It is clear that the limits become quite stringent away from the alignment limit.  For comparison, we include the
results of Ref.~\cite{Inoue:2014nva} for the limits from the electron EDM (eEDM, the blue shaded regions are excluded).  In all cases, the EDM limit and the heavy Higgs searches exclude complementary regions.  The masses of the heavy Higgs are increased to around 600 GeV in Fig.~\ref{fg:z2t1600}. In this case,
the limits from heavy Higgs searches become much weaker, with the dominant excluded region coming from the eEDM 
searches.   The mass splitting between the heavy masses is restricted by limits on the oblique parameters, which is discussed
in Sec. \ref{sec:stu}.

Fig.~\ref{fg:z2t2} shows the limits on $\alpha_b$ versus $\tan\beta$ in the Type-II model. 
Away from the alignment limit (the right panel), there is a significant exclusion region for $\tan\beta\gtrsim 10$ from the $126$ GeV
Higgs parameter measurements.  
Around $\tan\beta\sim 1$, the electron EDM constraint vanishes due to a cancellation 
among the Barr-Zee diagrams as pointed out in Ref.~\cite{Shu:2013uua}. 
We find the heavy Higgs searches from the  gauge boson decay channels 
$h_{2,3}\rightarrow Zh_1$ and $h_{2,3}\rightarrow WW/ZZ$ are
extremely useful and close the window of large values of $\sin\alpha_b\sim\mathcal{O}(1)$ in all cases.  
As the mass of the heavy particles is increased in Fig. \ref{fg:z2t2600}, the region excluded
by the heavy Higgs searches shrinks, with again the dominant exclusion coming from the eEDM and neutron EDM (nEDM, the green regions are excluded).
It is worth pointing out that the neutron EDM excluded regions are shown using the central values given in~\cite{Engel:2013lsa}, which however involves large uncertainties in the evaluation of hadronic matrix elements. 
In contrast, the heavy Higgs searches  provide a robust upper limit on the CP violating angle $\alpha_b$. 

\subsection{Limits from Heavy Higgs Searches in the Models with no $Z_2$ Symmetry}

As discussed in Sec.~\ref{sec:noz2}, if the assumption of an approximate $Z_2$ symmetry is relaxed, 
the theoretical relationship between $\alpha_b$ and $\alpha_c$ can be removed. In this case $\alpha_c$ becomes a free parameter.  
This helps to remove the theoretically inaccessible region in Figs.~\ref{z2t1}--\ref{fg:z2t2600}, 
and one can get a complete view of various constraints in the whole parameter space.

\begin{figure}[t]
\centerline{\includegraphics[width=0.9\columnwidth]{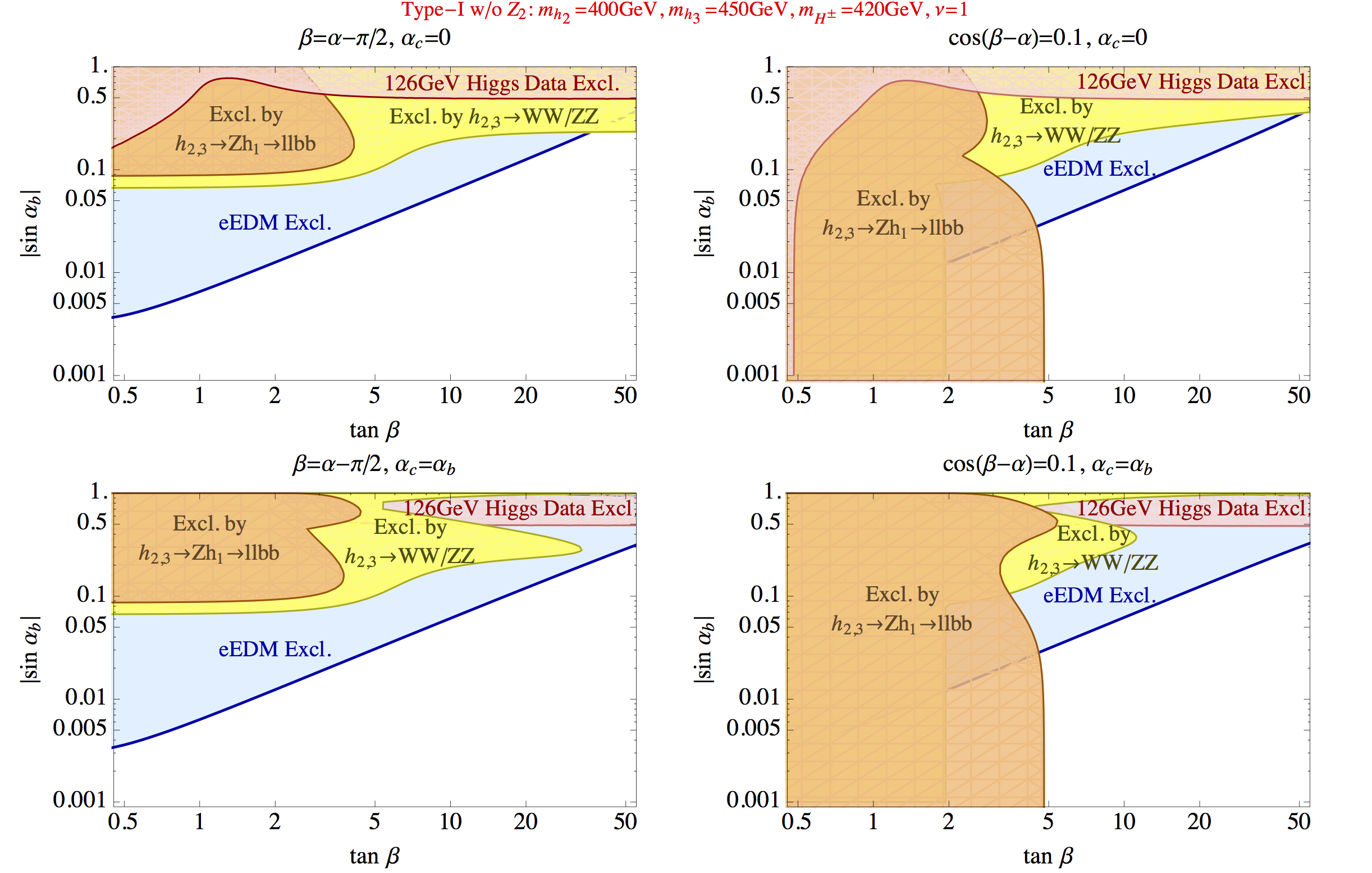}}
\caption{Heavy Higgs search constraints on the Type-I 2HDM without approximate $Z_2$ symmetry, {\it i.e.}, in this case $\alpha_c$ is a free parameter 
which is allowed to vary.
The color scheme for the exclusion regions is the same as in Figs.~\ref{z2t1}--\ref{fg:z2t2600}.
The first two rows use the same parameters as Fig.~\ref{z2t1}, and the last two rows use the same as Fig.~\ref{fg:z2t1600}.}
\label{fg:nz2t1}
\end{figure}

\begin{figure}[h]
\centerline{\includegraphics[width=0.9\columnwidth]{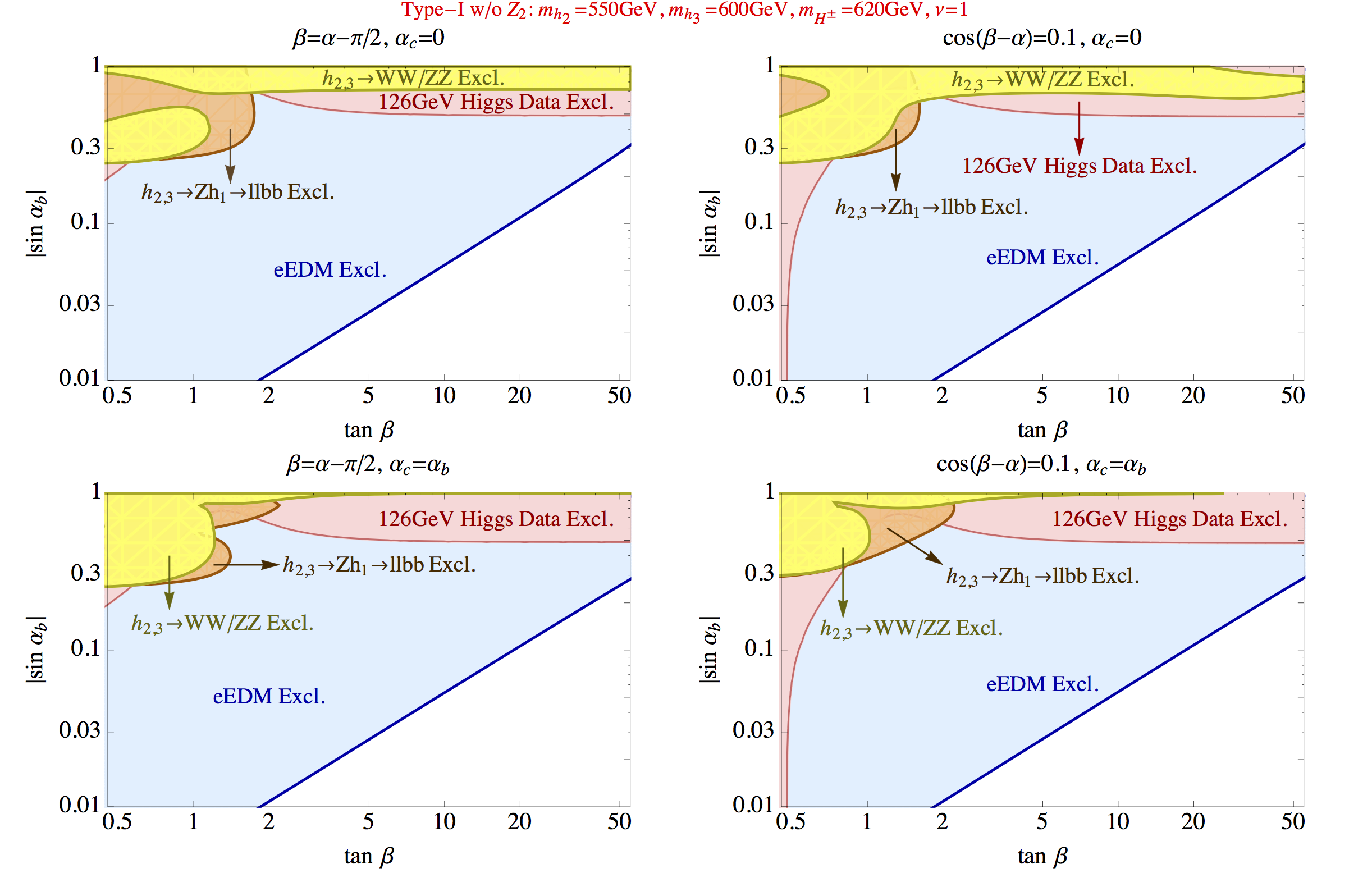}}
\caption{Similar to Fig.~\ref{fg:nz2t1}, but with heavy Higgs masses $m_{h_2}=550\,$GeV, $m_{h_3}=600\,$GeV, $m_{H^+}=620\,$GeV.\vspace{-2cm}}
\label{fg:nz2t1bbb}
\end{figure}

\begin{figure}[h]
\centerline{\includegraphics[width=0.9\columnwidth]{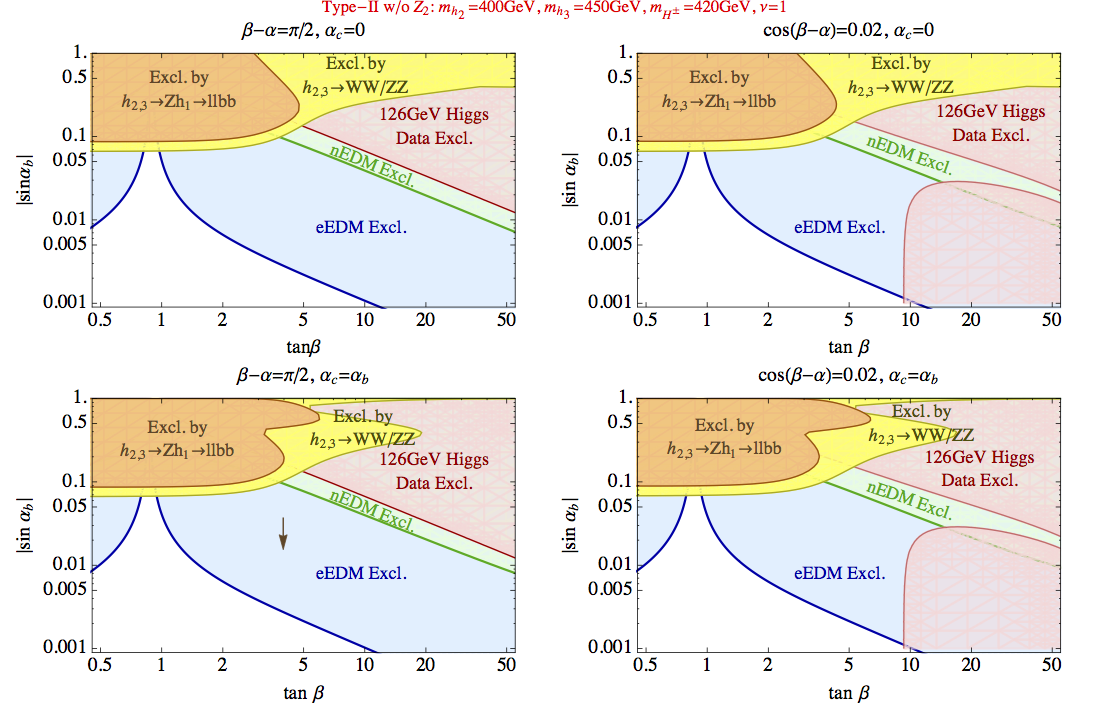}}
\caption{Similar to Fig.~\ref{fg:nz2t1}, but for the Type-II 2HDM without approximate $Z_2$ symmetry. }\label{fig8}
\end{figure}

\begin{figure}[h]
\centerline{\includegraphics[width=0.9\columnwidth]{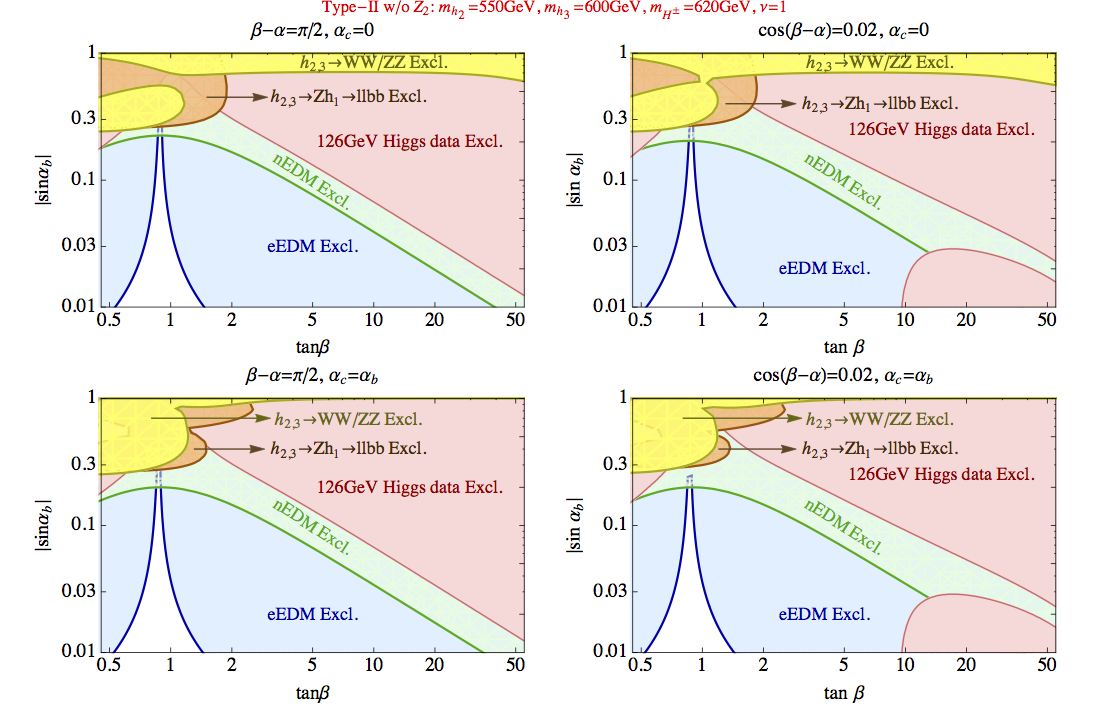}}
\caption{Similar to Fig.~\ref{fg:nz2t1bbb}, but for the Type-II 2HDM without approximate $Z_2$ symmetry.\vspace{-2cm}}
\label{fgnz2t2}
\end{figure}

\begin{figure}[t]
\centerline{\includegraphics[width=0.92\columnwidth]{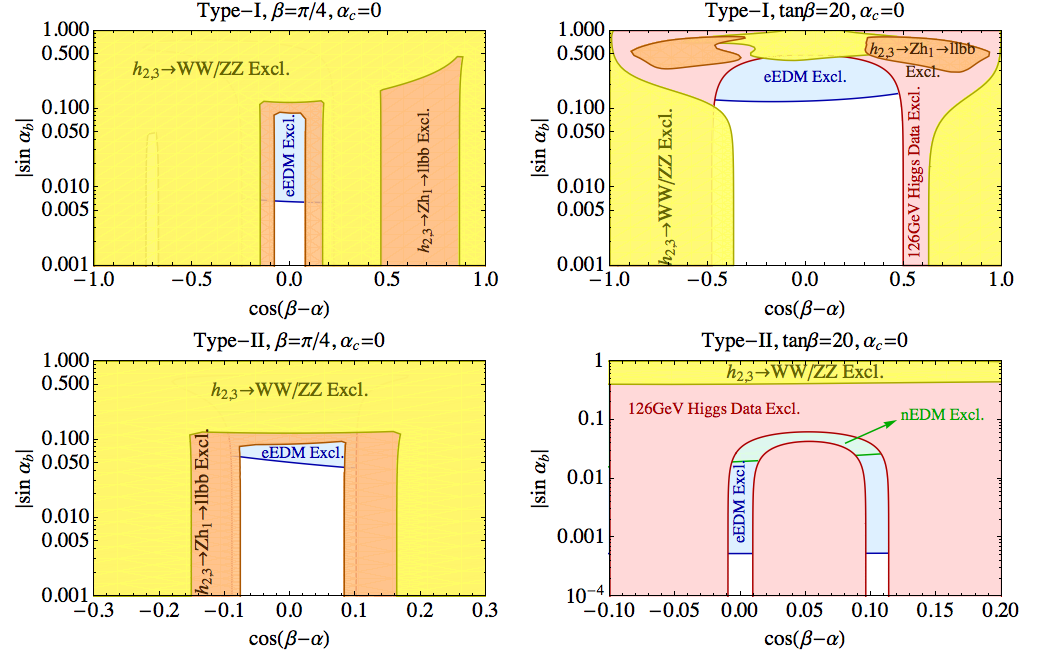}}
\caption{
Heavy Higgs search constraints on the Type-I (first row) and Type-II (second row) 2HDM without approximate $Z_2$ symmetry, 
using the $h_{2,3}\to WW/ZZ$ (yellow) and $h_{2,3}\to Z h_1\to l^+l^-b\bar b$ (orange) channels.
The heavy scalar masses are fixed to be $m_{h_2}=400\,$GeV, $m_{h_3}=450\,$GeV, $m_{H^+}=420\,$GeV, and the model parameter $\nu=1$. The other mixing angle $\alpha_c=0$.
}\label{bma1}
\end{figure}

\begin{figure}[t]
\centerline{\includegraphics[width=0.92\columnwidth]{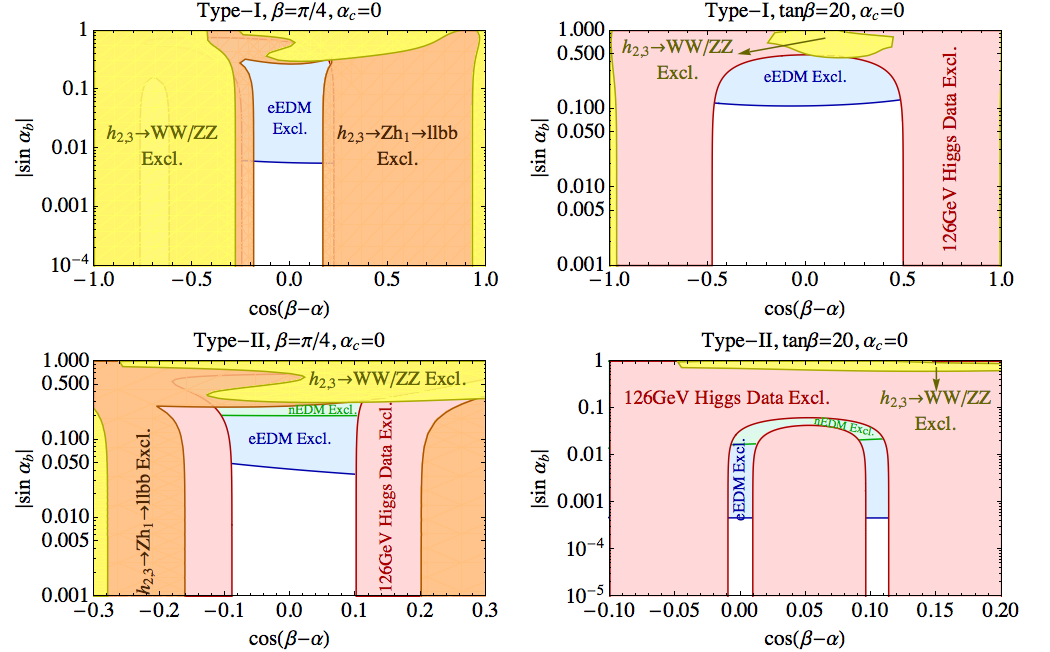}}
\caption{Similar to Fig.~\ref{bma1}, but with heavy scalar masses $m_{h_2}=550\,$GeV, $m_{h_3}=600\,$GeV, $m_{H^+}=620\,$GeV.\vspace{-2cm}}\label{bma2}
\end{figure}

Figs.~\ref{fg:nz2t1} and \ref{fg:nz2t1bbb} show the constraints in the Type-I model with $\alpha_c$ chosen equal to 0 or $\alpha_b$, and with two sets of heavy Higgs masses. It is apparent that the dependence on $\alpha_c$ is rather weak.  
The results in the Type-II model are shown in Figs.~\ref{fig8} and \ref{fgnz2t2}. 
In Figs.~\ref{bma1} and \ref{bma2},  the heavy Higgs search constraints are also displayed in the $\alpha_b$ and $\cos(\beta-\alpha)$ plane.

It is also worth re-emphasizing that at low $\tan\beta \sim \mathcal{O}(1)$ the 126 GeV Higgs data puts a very weak constraint on the CP violating angle $\alpha_b$.
This can also be understood from Table~\ref{Hcouplings}, where the lightest (126 GeV) Higgs couplings to other SM particles near the alignment limit are
\begin{align}\label{eq3.1}
& a_1 \simeq \cos\alpha_b \ , \ \ \  c_{t,1} \simeq (1+\Delta \cot\beta) \cos\alpha_b \ , \ \ \ \ \  \tilde c_{t,1} \simeq - \cot\beta \sin\alpha_b \ , \nonumber\\
& c_{b,1} \simeq \left\{\begin{array}{ll}
(1+\Delta \cot\beta) \cos\alpha_b, & \hspace{0.3cm} {\rm Type-I} \\
(1-\Delta \tan\beta) \cos\alpha_b, & \hspace{0.3cm} {\rm Type-II}
\end{array}
\right. \ \ \ \ \ \ \tilde c_{b,1} \simeq \left\{\begin{array}{ll}
\cot\beta \sin\alpha_b, & \hspace{0.3cm} {\rm Type-I} \\
-\tan\beta \sin\alpha_b, & \hspace{0.3cm} {\rm Type-II}
\end{array}
\right. 
\end{align}
where $\Delta=\cos(\beta-\alpha)$ and we have kept terms up to first power in $\Delta$.
Clearly for small $\Delta$ and $\tan\beta \approx 1$, all  CP even couplings are approximately $\cos\alpha_b$ and all CP odd couplings $\approx\pm\sin\alpha_b$. They approach  the values in the SM limit when $\alpha_b\to0$.
In the presence of CP violation, the gluon fusion production cross section of the light Higgs gets rescaled from the SM value by a factor~\cite{Shu:2013uua, Inoue:2014nva}, $\sigma(gg\to h_1)/\sigma_{\rm SM}(gg\to H_{SM}) \simeq 1 + 1.42 \sin^2\alpha_b$.
The vector boson fusion and associated production rates get suppressed by $\sigma(VV\to h_1)/\sigma_{\rm SM}(VV\to H_{SM}) = \cos^2\alpha_b$.
The light Higgs to fermion ($h_1\to b\bar b, \tau^+\tau^-$) decay rates are not affected because the CP even and CP odd couplings contribute incoherently, $\Gamma(h_1\to f\bar f)/\Gamma_{\rm SM}(h\to f\bar f) = \cos^2\alpha_b+\sin^2\alpha_b =1$. The Higgs to gauge boson ($h_1\to WW^*, ZZ^*$) decay rates get suppressed, $\Gamma(h_1\to VV^*)/\Gamma_{\rm SM}(h_1\to VV^*) = \cos^2\alpha_b$.
The light Higgs to diphoton decay rate in the presence of CP violation has been given in refs.~\cite{Shu:2013uua, Inoue:2014nva}, which in this case can be simplified to $\Gamma(h_1\to\gamma\gamma)/\Gamma_{\rm SM}(h\to\gamma\gamma) \simeq 1-0.81 \sin^2\alpha_b$.
As a result, the final $\chi^2$ of the fit for the 126 GeV Higgs data depends on $\cos^2\alpha_b$, and for the SM case $\chi^2_{\rm SM} = \chi^2(\cos^2\alpha_b\to 1)$.
Because the $\cos^2\alpha_b$ function is very flat near $\alpha_b=0$, one can maintain a fit as good as in the SM for sizable $\alpha_b$.

In contrast, the heavy Higgs decay to gauge boson channels ($h_{2,3}\to VV$ and $Zh_1$) are more sensitive to a 
non-zero CP violating angle $\alpha_b$ and can place a stronger constraint on it. This feature has been discussed in Sec.~\ref{section2.6}. Furthermore, from the figures we notice 
that at low $\tan\beta$, the heavy Higgs search constraint is stronger than at large $\tan\beta$.
This is because $h_i\to t\bar t, (i=2,3)$ is the dominant decay mode and the branching ratio for the gauge boson decay modes of $h_i$ can be written as
\begin{eqnarray}
{\rm Br}_{h_i \to VV\, {\rm or}\, Zh} ({\rm low}\ \tan\beta) \sim \frac{\Gamma_{h_i \to VV\, {\rm or}\, Zh}}{\Gamma_{h_i\to t\bar t} } \ .
\end{eqnarray}
Eqs.~(\ref{Csample}) and (\ref{Asample}) tell us  that these two rates around the alignment limit are both insensitiveto  variations  of $\tan\beta$.
However, as $\tan\beta$ grows to larger than ${\cal{O}}(2)$,  the other decay channels such as $h_i\to h_1 h_2$ and $h_i\to b\bar b$ larger than 
 $h_i\rightarrow t\bar t$, and they are not yet constrained by the LHC data. As a result, the gauge boson decay rates of heavy Higgs bosons are 
 suppressed in this region.

\begin{figure}[t]
\centerline{\includegraphics[width=0.78\columnwidth]{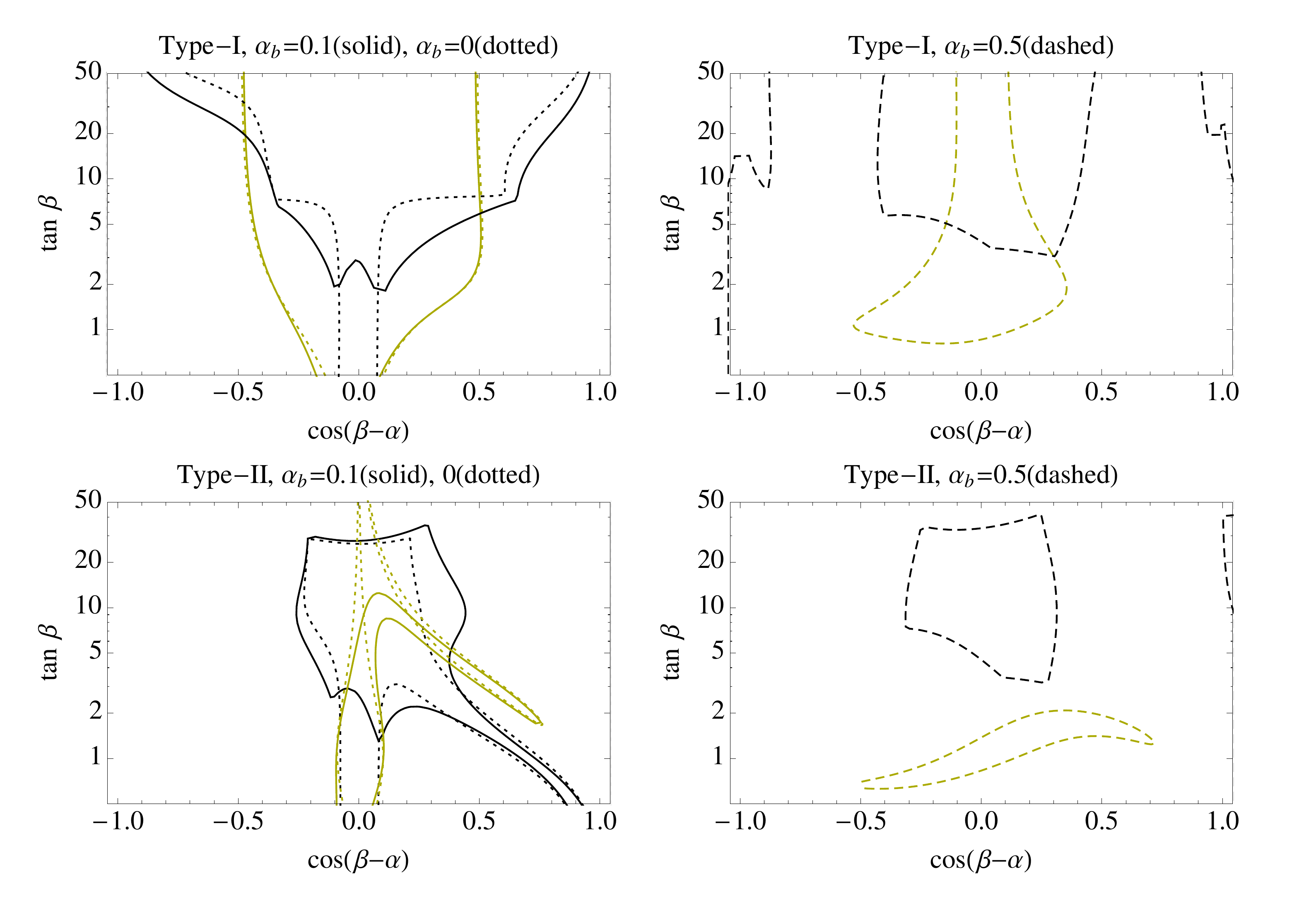}}
\caption{Heavy Higgs search (black) and 126 GeV Higgs data (yellow) constraints at 95\% CL on the Type-I (first row) and Type-II (second row) 2HDM without approximate $Z_2$ symmetry. Different curves correspond to $\alpha_b =0$ (dotted), 0.1 (solid) and 0.5 (dashed).
The heavy Higgs curves include the combination of constraints from $h_{2,3}\to WW/ZZ$, $h_{2,3}\to Z h_1\to l^+l^-b\bar b$ and $h_{2,3}\to \tau^+ \tau^-$ channels.
The heavy scalar masses are fixed to be $m_{h_2}=400\,$GeV, $m_{h_3}=450\,$GeV, $m_{H^\pm}=420\,$GeV, and the model parameter $\nu=1$. The other mixing angle $\alpha_c=0$.}\label{tanbcosbma}
\end{figure}

\begin{figure}[t]
\centerline{\includegraphics[width=0.78\columnwidth]{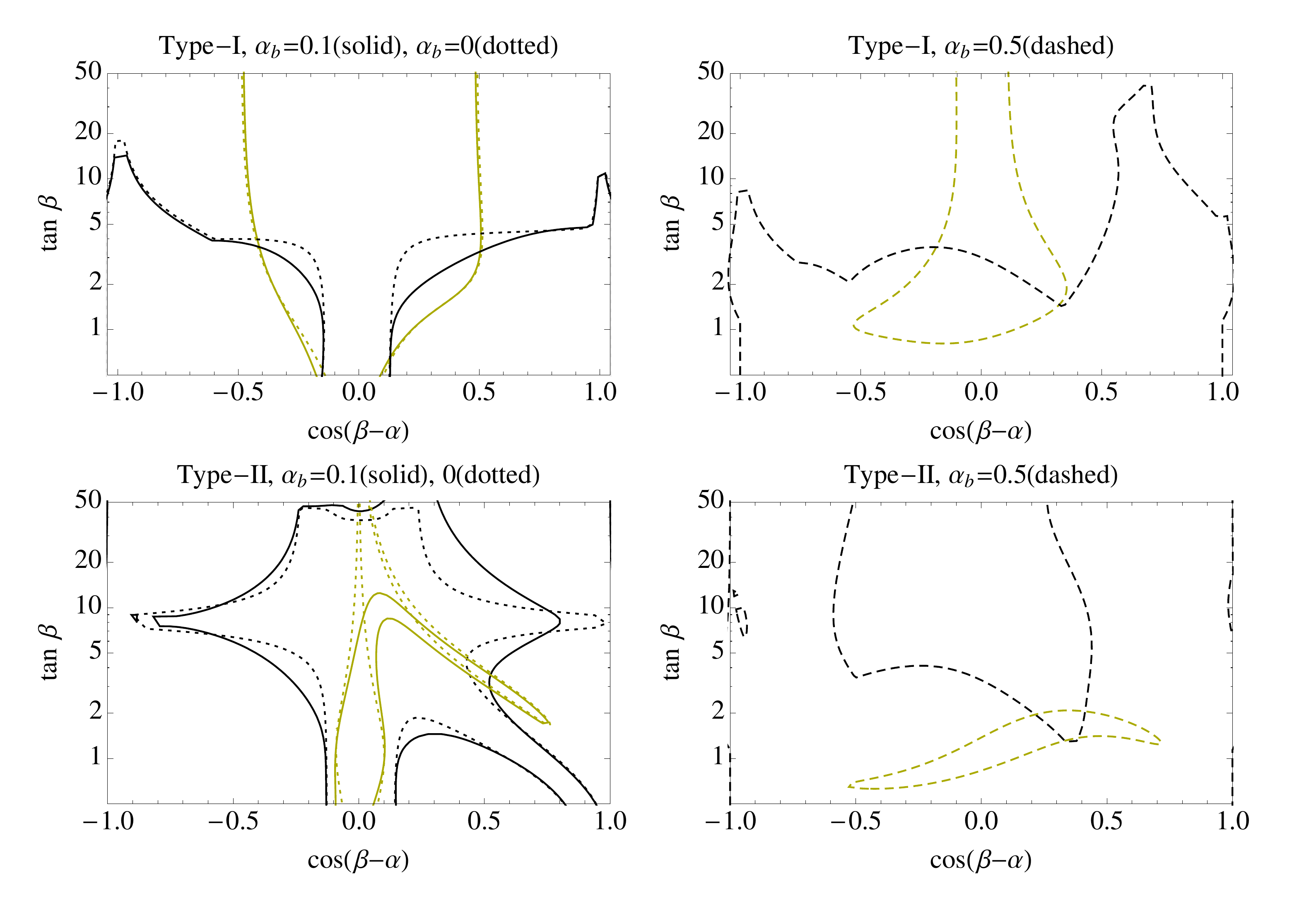}}
\caption{Similar to Fig.~\ref{tanbcosbma}, but for heavy Higgs masses $m_{h_2}=550\,$GeV, $m_{h_3}=600\,$GeV, $m_{H^\pm}=620\,$GeV.\vspace{-2cm}}
\label{tanbcosbma2}
\end{figure}

Fig.~\ref{tanbcosbma} depicts 95\% CL constraints in the $\tan\beta$ versus $\cos(\beta-\alpha)$ plane from heavy Higgs searches (black) and 
from 126 GeV Higgs data (yellow) on the Type-I (first row) and Type-II (second row) 2HDMs without approximate $Z_2$ symmetry. Different curves correspond to $\alpha_b =0$ (dotted), 0.1 (solid) and 0.5 (dashed), and the other mixing angle $\alpha_c=0$. For the CP conserving case $(\alpha_b = 0)$, we found that the bounds are very similar to those studied in Refs. \cite{Chen:2013kt,Chen:2013rba,Craig:2013hca,Chen:2013qda}. 
In both Type-I and Type-II models, both heavy and light Higgs searches favor regions around the alignment limit $\cos(\beta-\alpha)=0$. 
In the Type-II model when CP violation is small (bottom left panel), there is another allowed branch corresponding to $\cos(\beta+\alpha)\sim0$~\cite{Ferreira:2014naa}, but we find the heavy and light Higgs favored regions are inconsistent with each other for very large deviations from the  alignment limit.
In the Type-I model (first row), the light Higgs bound only depends on $\cos(\beta-\alpha)$ , but is independent of $\tan\beta$ in the large $\tan\beta$ limit. The reason is that in this case the $h_1$ couplings can be approximated as $c_{t,1}=c_{b,1} \to \sin(\beta-\alpha) \cos\alpha_b + \mathcal{O}(1/\tan\beta)$, $\tilde c_{t,1} = -\tilde c_{b,1} = \mathcal{O}(1/\tan\beta)$ and $a_1=\sin(\beta-\alpha)\cos\alpha_b$, so their dependence on $\tan\beta$ is suppressed.
On the other hand, for the Type-II model, the couplings ${c}_{b,1}, {c}_{\tau,1}$ and $\tilde{c}_{b,1}, \tilde{c}_{\tau,1}$ are enhanced at large $\tan\beta$.
This explains why in the Type-II model (second row), light Higgs data are more restrictive on the parameter space with large $\tan\beta$. 
As a result, for $\alpha_b=0.5$,  the light Higgs data only favors a region with $\tan\beta\lesssim2$ (see the bottom right panel of Fig.~\ref{tanbcosbma}). In contrast, we have learned that the heavy Higgs search data are more sensitive at small $\tan\beta$ and for $\alpha_b=0.5$ they only allow the region where $\tan\beta\gtrsim3$, thus there is no region in the parameter space that can be made consistent with both light and heavy Higgs results from LHC.
Fig.~\ref{tanbcosbma2} gives results similar to those in Fig. \ref{tanbcosbma} but with a different set of mass parameters, $m_{h_2}=550\,$GeV, $m_{h_3}=600\,$GeV, $m_{H^+}=620\,$GeV. The parameter space becomes less constrained by the heavy Higgs searches because the production cross sections  are smaller compared to those in Fig.~\ref{tanbcosbma}.

From the above results, we can conclude that if the heavy Higgs masses lie below around 600 GeV, 
the CP violating phase $\alpha_b$ is constrained to be less than around 30\% throughout the most general parameter space.
The regions which allow  $\alpha_b$ close to this upper bound are $\tan\beta\sim1$ in the Type-II model, and
$\tan\beta\gtrsim20$ in the Type-I model without an approximate $Z_2$ symmetry.
We have also estimated the future sensitivity of the heavy Higgs search at the 
14 TeV LHC by rescaling the current limits by the square root of expected number of events ($\sigma \times \mathcal{L}$).
With 300 (3000) fb$^{-1}$ data, if the heavy Higgs masses are below 600 GeV and we still do not find them, the CP violating angle $\alpha_b$ will be constrained to be less than around $10$\%.

Recall that the angle $\alpha_b$ parametrizes the size of CP odd mixture in the 126 GeV Higgs boson. 
The main point of this work is to show that the heavy Higgs search is relevant and plays a complimentary role to the other indirect searches, 
and sometimes it stands at the frontier of probing the Higgs boson CP mixture.

\newpage

\newpage

$ $

\newpage

\section{Limits from $B$ Decays, Oblique Parameters, and $(g-2)_\mu$}
\label{explims}
The CP violating 2HDM is also  limited by measurements in $B$ decays, the oblique parameters, and $(g-2)_\mu$.  In Type-II
models the charged Higgs mass is restricted by $B$ data to be greater than $m_{H^+}\sim 340\,{\rm GeV}$ for all values of $\tan\beta$.
In both Type-1 and Type-2 models, measurements in the $B$ system prefer $\tan\beta >1$~\cite{Mahmoudi:2009zx,Chen:2013kt, WahabElKaffas:2007xd}. 

\subsection{Limits from Electroweak Oblique Parameters}
The allowed parameters are restricted by measurements of the oblique parameters.  The general results for $S,T$ and $U$ in
a 2HDM are given in Refs.~\cite{Grimus:2008nb,He:2001tp,Branco:2011iw, Haber:2010bw}.
In the alignment limit, $\cos\alpha=\sin\beta$ and $\sin\alpha=-\cos\beta$, the results simplify considerably,
\begin{eqnarray}
\alpha \Delta T&=&{1\over 16 \pi^2 v^2}\biggl\{
\sin^2 \alpha_b F(m_{H^+}^2, m_{h_1}^2)
+(1-\sin^2\alpha_b \sin^2\alpha_c)F(m_{{H^+}}^2,m_{h_2}^2)\nonumber \\ 
&&\hspace{1.6cm}+(1-\sin^2\alpha_b\cos^2\alpha_c)F(m_{H^+}^2,m_{h_3}^2)
-\cos^2\alpha_c\sin^2\alpha_bF(m_{h_1}^2,m_{h_2}^2)\nonumber \\
&&\hspace{1.6cm}-\sin^2\alpha_c\sin^2\alpha_bF(m_{h_1}^2,m_{h_3}^2)
-\cos^2\alpha_bF(m_{h_2}^2,m_{h_3}^2)
\nonumber \\ 
&&\hspace{1.6cm}+3\cos^2\alpha_b\biggl[F(M_Z^2,m_{h_1}^2)-F(M_W^2,m_{h_1}^2)\biggr]\nonumber \\
&&\hspace{1.6cm}+3\sin^2\alpha_c\sin^2\alpha_b\biggl[F(M_Z^2,m_{h_2}^2)-F(M_W^2,m_{h_2}^2)\biggr]
\nonumber \\ 
&&\hspace{1.6cm}+3\cos^2\alpha_c\sin^2\alpha_b[F(M_Z^2,m_{h_3}^2)-F(M_W^2,m_{h_3}^2)\biggr]
\nonumber \\
&&\hspace{1.6cm}-3\biggl[F(M_Z^2, M_{H,ref}^2)-F(M_W^2,M_{H,ref}^2)\biggr]
\biggr\}\, ,
\label{tans}
\end{eqnarray}
where the last line is the subtraction of the SM Higgs contribution evaluated at the reference scale, $M_{H,ref}$, at which the 
fit to the data is performed.   The function $F(x,y)$ is, 
\begin{eqnarray}
F(x,y)&=&{x+y\over 2}-{xy\over (x-y)}\log\biggl({x\over y}\biggr)\ .
\nonumber \\
F(x,x)&=& 0 \ ,\nonumber \\
F(x,y)&\xrightarrow{y\gg x}&{y\over 2}\ .
\end{eqnarray}
With $\alpha_c$=0, we obtain  the simple form,
\begin{eqnarray}
\alpha \Delta T&=& {1\over 12\pi ^2 v^2}\biggl\{
\Delta_2\Delta_3\cos^2 \alpha_b+\biggl[
\Delta_1\Delta_2-2(\Delta_3-\Delta_1)(M_W-M_Z)\biggr]\sin^2\alpha_b\biggr\}
\label{tsim}
\end{eqnarray}
and  $\Delta_i\equiv m_{H^+}-m_{h_i}$.
Eq.~(\ref{tsim}) is in agreement with Ref.~\cite{Barbieri:2006dq} in the limit $\alpha_b=0$.

\begin{figure}[t]
\centerline{\includegraphics[width=1.0\columnwidth]{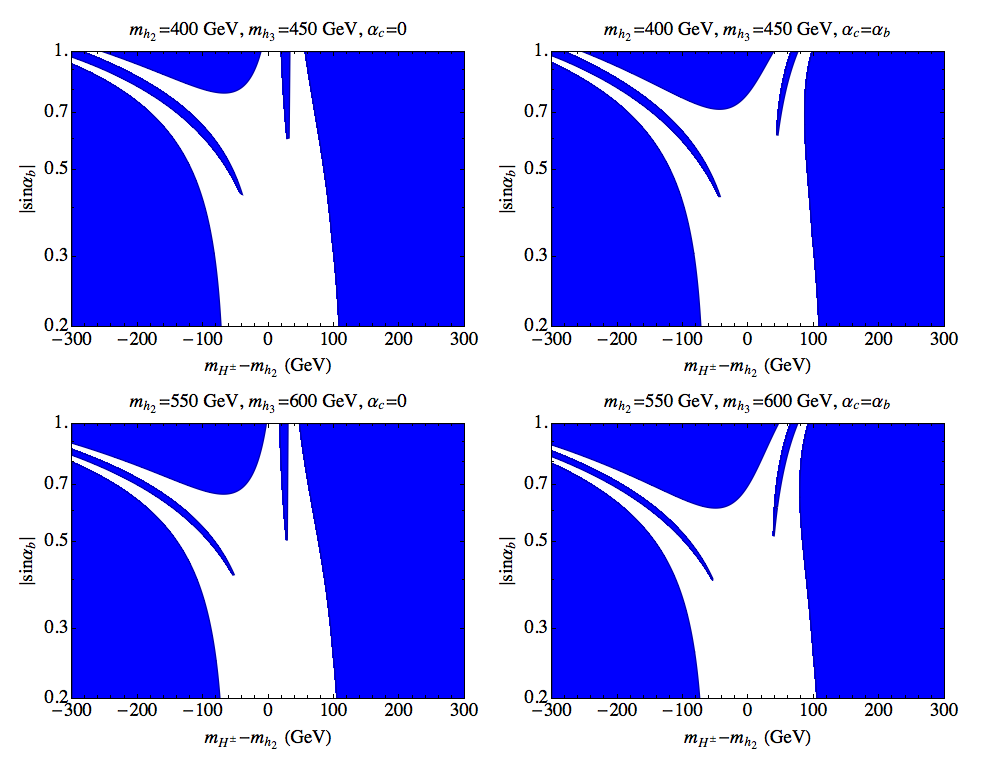}}
\caption{$95\%$ Confidence Level allowed regions (white) from fits to the oblique parameters in the CP violating 2HDM. }
\label{fg:stu}
\end{figure}

The result for $\Delta S$ also takes a simple form in the alignment limit~\cite{Grimus:2008nb},
\begin{eqnarray}
\Delta S&=&{1\over 24\pi}\biggl\{\cos^22\theta_WG(m_{H^+}^2,m_{H^+}^2, M_Z^2)+\sin^2\alpha_b
\biggl[\cos^2\alpha_c G(m_{h_1}^2,m_{h_2}^2,M_Z^2)
\nonumber \\ 
&&\hspace{1.2cm}+\sin^2\alpha_c G(m_{h_1}^2,m_{h_3}^2,M_Z^2)
+\sin^2\alpha_c {\hat G}(m_{h_2}^2, M_Z^2)
+\cos^2\alpha_c{\hat G}(m_{h_3}^2, M_Z^2)
\biggr]\nonumber \\ 
&&\hspace{1.2cm}
+\cos^2\alpha_b \biggl[ {\hat G}(m_{h_1}^2,M_Z^2) + G(m_{h_2}^2, m_{h_3}^2, M_Z^2) \biggr] 
+\ln\biggl({m_{h_1}^2 m_{h_2}^2 m_{h_3}^2\over m_{H^+}^6}\biggr)
\nonumber \\
&&\hspace{1.2cm}
-\biggl[{\hat G}(M_{H,ref}^2,M_Z^2)+\ln\biggl({M_{H,ref}^2\over m_{H^+}^2}\biggr)\biggr]
\biggr\}\, .
\label{sans}
\end{eqnarray}  Analytic results for $G(x,y,z)$ and ${\hat G}(x,y)$
are given in the appendix of Ref.~\cite{Grimus:2008nb}.

We use the Gfitter fit to the electroweak data~\cite{Baak:2014ora},
\begin{eqnarray}
S&=&0.05\pm 0.11\nonumber \\
T&=&0.09\pm 0.13\nonumber\\
U&=& 0.01\pm 0.11\, ,
\label{fit}
\end{eqnarray}
with a reference value for the SM Higgs mass, $M_{H,ref}=125~GeV$.  The $STU$ correlation matrix is,
\begin{eqnarray} \rho_{ij}&=&
\begin{pmatrix}
1& 0.90 & -0.59\\
0.09& 1& -0.83\\
-0.59 & -0.83 & 1
\end{pmatrix}\, ,
\end{eqnarray} 
and the $\chi^2$  is defined as
\begin{equation}
\Delta\chi^2=\Sigma_{ij}(\Delta X_i-\Delta {\hat X}_i)(\sigma^2)_{ij}^{-1}
(\Delta X_j-\Delta {\hat X}_j)\, ,
\end{equation}
where ${\hat X}_i=\Delta S, \Delta T$, and $\Delta U$ are the central values of the fit in Eq.~(\ref{fit}),
${\hat X}_i=\Delta S, \Delta T$, and $\Delta U$ are the parameters in the 2HDM (Eqs.~(\ref{tans}) and (\ref{sans})),
$\sigma_i$ are the errors given in Eq.~(\ref{fit}) and $\sigma_{ij}^2=\sigma_i\rho_{ij} \sigma_j$.  

In Fig.~\ref{fg:stu} we show the $95\%$ confidence level allowed regions  for $\alpha_b=\alpha_c$
and $\alpha_c=0$.  For $\alpha_b$ close to $1$, there is some interesting structure due to the interplay of the
$\Delta S$ and $\Delta T$ limits.  For $\mid \sin\alpha_b \mid<  0.5$, the results are well approximated by the limit
from $\Delta T$ only,
\begin{equation}
-80~{\rm GeV} <  \Delta_2 < 120~{\rm GeV}
\, .
\end{equation}

\label{sec:stu}

\subsection{Limits from muon $g-2$}
The experimentally measured value of ${(g-2)_\mu\over 2}=a_\mu$ places a weak constraint on the parameters of the CP violating 2HDM.  
The deviation between the experimental number and the SM theory prediction is~\cite{Davier:2004gb},
\begin{equation}
\Delta a_\mu=a_\mu^{exp}-a_\mu^{SM}=265 (85)\times 10^{-11}\, .
\label{gm2exp}
\end{equation}

The one-loop contributions from the Higgs sector in the 2HDM to $\Delta a_\mu$  
are numerically small.  The larger Higgs sector contributions come from the 2-loop Barr-Zee type diagrams with a closed fermion/gauge-boson/heavy-Higgs
loop.  This class of diagrams can be enhanced by factors of $M^2/m_\mu^2$ relative to the 1-loop diagrams, where $M$ is a heavy
Higgs or heavy fermion mass.  For completeness,
these results are  given in Appendix C.  

In Figs.~\ref{fg:gm2_1} and \ref{fg:gm2_2}, we show the contributions to $\Delta a_\mu$ in the 2HDM for relatively heavy $m_{2,3}$ and $m_{H^+}$
in units of $10^{-11}$.  For $\mid \sin\alpha_b\mid\lesssim 0.5$, there is almost no sensitivity to the CP violating phase.  The largest contribution is
found in the Type-II model for large $\tan\beta$ and is of opposite sign to that needed to explain the discrepancy of Eq.~(\ref{gm2exp}).   

\begin{figure}[t]
\centerline{\includegraphics[width=1.0\columnwidth]{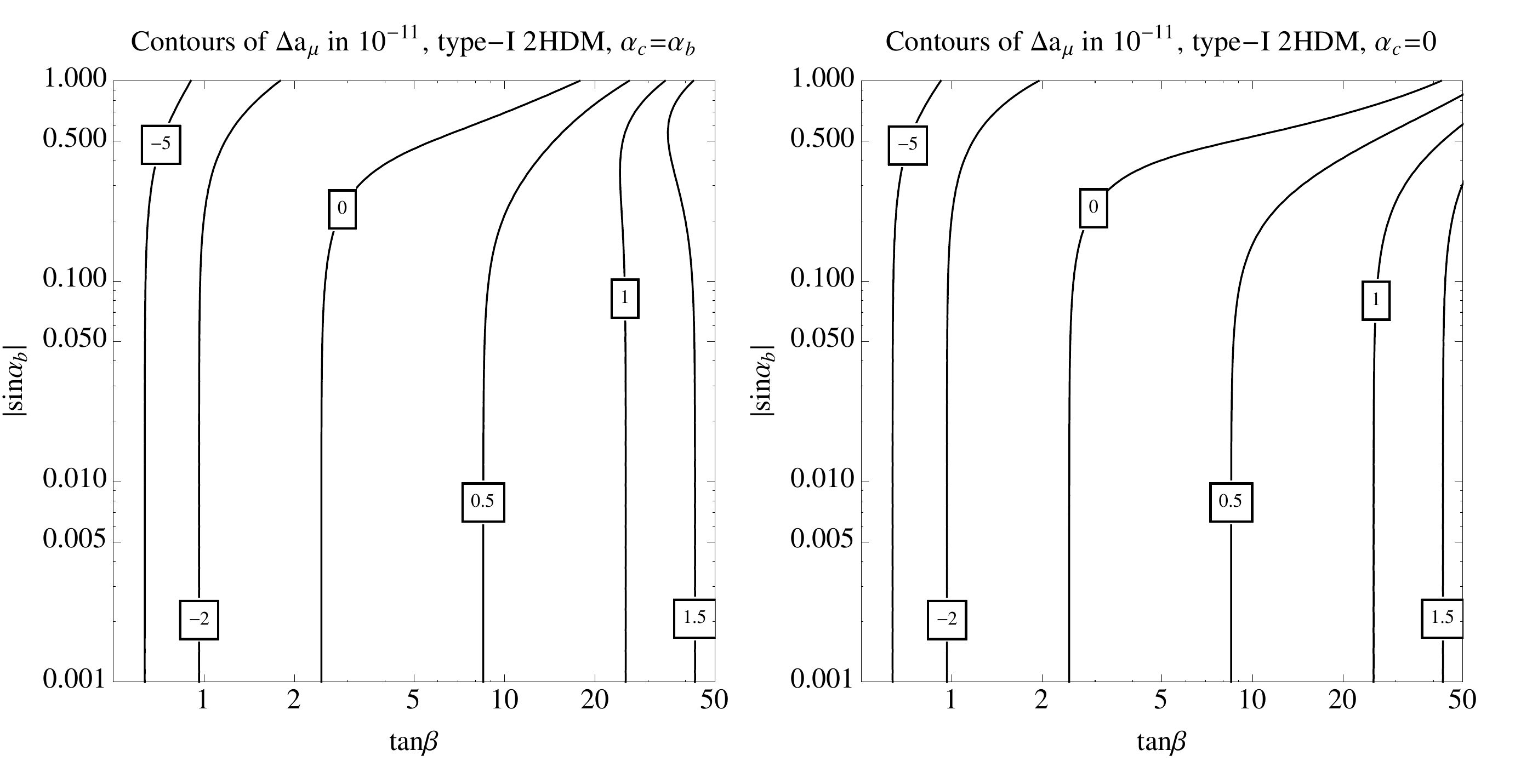}}
\caption{Contributions to $(g-2)_\mu$ in the CP violating Type-I 2HDM from the Barr-Zee diagrams. The heavy scalar masses are fixed to be $m_{h_2}=400\,$GeV, $m_{h_3}=450\,$GeV, $m_{H^+}=420\,$GeV, and the model parameter $\nu=1$.}
\label{fg:gm2_1}
\end{figure}

\begin{figure}[t]
\centerline{\includegraphics[width=1.0\columnwidth]{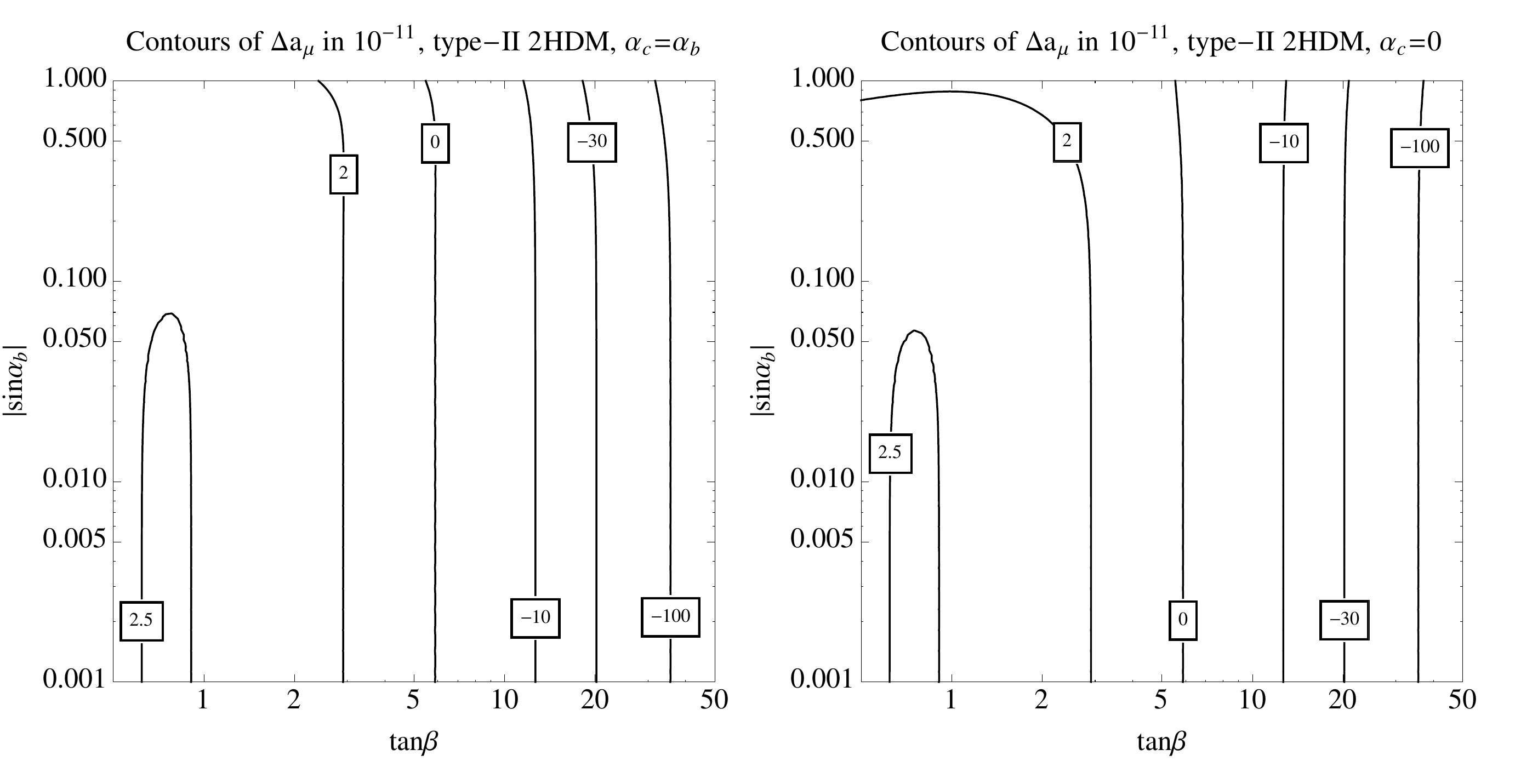}}
\caption{Similar to Fig.~\ref{fg:gm2_1} but for Type-II 2HDM.}
\label{fg:gm2_2}
\end{figure}

\label{sec:gm2}

\section{Conclusion}
The CP mixture of the 126 GeV Higgs boson is an important property  of the Higgs sector that deserves further scrutiny.  
A non-zero CP component is theoretically well motivated and may be the origin of the  cosmic baryon asymmetry.
An important consequence of the 126 GeV Higgs boson having a sizable CP odd mixture is that the new physics responsible for this cannot be decoupled and must lie near the electroweak scale.

In the context of CP violating, flavor conserving two-Higgs-doublet models, we studied the impact of the heavy Higgs searches at
the  LHC on the CP violating parameters.
In this class of models, CP violation appears in the neutral Higgs sector, where there are two more real scalars ($h_{2,3}$) 
in addition to  the lightest 126 GeV one.
The couplings of the heavy Higgs scalars with electroweak gauge bosons are very sensitive to the CP violation in the Higgs sector. Turning on a CP odd mixture in the 126 GeV Higgs boson will also turn on the heavy Higgs decay channels into gauge bosons, $h_{2,3}\to WW/ZZ$ and $Z h_1$. 
There is data from the LHC from  the search for a SM like Higgs boson in these decay channels, and the non-discovery of a heavy Higgs can be re-interpreted as constraints on the allowed deviation from the alignment limit in the two-Higgs-doublet models without CP violation. 

In this work, we point out that heavy Higgs searches are  also extremely useful for  constraining Higgs sector CP violation and in particular the CP mixture of the 126 GeV Higgs boson.
We demonstrate that the constraints from heavy Higgs searches are largely complimentary to the low energy EDM constraints.
We compare our results with the limits from the  global fit to the 126 GeV Higgs data, and find they can place much stronger limits than the 
light Higgs coupling fit, especially in the  
interesting regions when there are destructive contributions to the EDM.
We find in these regions that the heavy Higgs searches are at the frontier of probing Higgs sector CP violation.
The current limit on the CP violating mixing angle, parametrized by $\alpha_b$, is constrained to be less than 30\%, and the LHC heavy Higgs search can further narrow down the angle to less than  a 10\% level with the high luminosity runs.
We also expect our work to be a roadmap for the future searches for Higgs sector CP violation and the exciting interplay across various experimental frontiers.

For completeness, we have also explored other relevant constraints from electroweak oblique parameters, the muon $g-2$ and 
from B physics, and discussed their implications on the heavy Higgs parameter limits  in CP violating 2HDMs.

\label{sec:conc}

\appendix
\numberwithin{equation}{section}
\section{Solving the potential parameters in the approximate $Z_2$ case}
\label{appa}

In this section, we list the relations between the potential parameters and the phenomenological parameters listed in Eq.~(\ref{sec:2.3}) in the approximate $Z_2$ symmetric 2HDMs.
\begin{eqnarray}\label{mini}
m_{11}^2 &=& \lambda_1 v^2 \cos^2\beta + (\lambda_3 + \lambda_4) v^2 \sin^2\beta - {\rm Re} (m_{12}^2 e^{i\xi}) \tan\beta + {\rm Re} (\lambda_5 e^{2i\xi}) v^2\sin^2\beta \ , \hspace{0.8cm}
\label{mini1}\\
m_{22}^2 &=& \lambda_2 v^2 \sin^2\beta + (\lambda_3 + \lambda_4) v^2 \cos^2\beta - {\rm Re} (m_{12}^2 e^{i\xi}) \cot\beta + {\rm Re} (\lambda_5 e^{2i\xi}) v^2\cos^2\beta \ ,
\label{mini2}\\
{\rm Im} (m_{12}^2)&=&v^2 \sin\beta\cos\beta {\rm Im} ( \lambda_5 ) \ , \\
\lambda_1 &=& \frac{m_{h_1}^2 \sin^2\alpha \cos^2\alpha_b + m_{h_2}^2 R_{21}^2 
+ m_{h_3}^2 R_{31}^2}{v^2 \cos\beta^2} - \nu \tan^2\beta \ , \\
\lambda_2 &=& \frac{m_{h_1}^2 \cos^2\alpha \cos^2\alpha_b + m_{h_2}^2 R_{22}^2 
+ m_{h_3}^2 R_{32}^2}{v^2 \sin\beta^2} - \nu \cot^2\beta \ , \\
\lambda_4 &=& 2 \nu - {\rm Re}\lambda_5 - \frac{2 m_{H^+}^2}{v^2} \ , \\
\lambda_3 &=& \nu - \frac{m_{h_1}^2 \sin\alpha \cos\alpha \cos^2\alpha_b - m_{h_2}^2R_{21}R_{22} - m_{h_3}^2R_{31}R_{32}}{v^2\sin\beta\cos\beta} - \lambda_4 - {\rm Re}\lambda_5 \ , \\
{\rm Re}\lambda_5 &=& \nu - \frac{m_{h_1}^2 \sin^2\alpha_b + \cos^2\alpha_b (m_{h_2}^2 \sin^2\alpha_c + m_{h_3}^2 \cos^2\alpha_c)}{v^2} \ , \\
{\rm Im}\lambda_5 &=& \frac{2 \cos\alpha_b}{v^2 \sin\beta} 
\left[ (m_{h_2}^2-m_{h_3}^2) \cos\alpha \sin\alpha_c \cos\alpha_c   \rule{0mm}{4.5mm}\right. \nonumber \\
&&\hspace{3.2cm}
\left. +(m_{h_1}^2 - m_{h_2}^2 \sin^2\alpha_c-m_{h_3}^2\cos^2\alpha_c) \sin\alpha \sin\alpha_b \rule{0mm}{4.5mm}\right] \ . \label{eq:imlambda5} 
\end{eqnarray}
There is an additional constraint,
\begin{eqnarray}
\tan\beta &=& \frac{(m_{h_2}^2 -m_{h_3}^2) \cos\alpha_c \sin\alpha_c + (m_{h_1}^2 -m_{h_2}^2 \sin^2\alpha_c-m_{h_3}^2 \cos^2\alpha_c) \tan\alpha \sin\alpha_b}
{(m_{h_2}^2 -m_{h_3}^2) \tan\alpha \cos\alpha_c \sin\alpha_c - (m_{h_1}^2 -m_{h_2}^2 \sin^2\alpha_c-m_{h_3}^2 \cos^2\alpha_c) \sin\alpha_b} \ . \label{ab}\nonumber \\
\end{eqnarray}

\section{Tri-linear Higgs Couplings}
From the quartic terms in the scalar potential Eq.~(\ref{pot}), we can obtain the interactions between three neutral scalars, in the basis of $(H_1^0, H_2^0, A^0)$,
\begin{scriptsize}
\begin{eqnarray}
\mathcal{L}_{3s} &=&\frac{1}{4} (A^0)^3 \cos\beta \left\{2\sin\beta {\rm Im}\lambda_5 - \cos\beta {\rm Im}\lambda_7 \rule{0mm}{5mm}\right\} \nonumber \\
&+& \frac{1}{8} (A^0)^2 \left\{ \left[ -5 H_1^0 \cos\beta + H_1^0 \cos(3\beta) - H_2^0 \left( 5 \sin\beta + \sin(3\beta) \right) \rule{0mm}{4mm}\right] {\rm Re}\lambda_5 \rule{0mm}{5mm}\right. \nonumber \\
&& \hspace{1.4cm}\left.+ 4 \left[ H_1^0 \cos\beta \sin^2\beta \lambda_1 + H_2^0 \cos^2\beta \sin\beta \lambda_2 + \left( H_1^0 \cos^3\beta + H_2^0 \sin^3\beta \right)(\lambda_3 + \lambda_4) \rule{0mm}{4mm}\right] \rule{0mm}{5mm}\right\} \nonumber \\
&+& \frac{1}{4} A^0 \left\{ \left[ 4 H_1^0 H_2^0 + \left((H_1^0)^2+(H_2^0)^2\right)\sin(2\beta) \rule{0mm}{4mm}\right]{\rm Im}\lambda_5 
+ H_2^0 \left( 2 H_2^0 - H_2^0 \cos(2\beta) + H_1^0 \sin(2\beta) \right) {\rm Im}\lambda_7 \rule{0mm}{5mm}\right\} \nonumber \\
&+& \frac{1}{2} \left\{ H_2^0 \sin\beta \left[ (H_2^0)^2 \lambda_2 + (H_1^0)^2 \left( \lambda_3 + \lambda_4 + {\rm Re}\lambda_5 \right) \rule{0mm}{4mm}\right]
+ H_1^0 \cos\beta \left[ (H_1^0)^2 \lambda_1 + (H_2^0)^2 \left( \lambda_3 + \lambda_4 + {\rm Re}\lambda_5 \right) \rule{0mm}{4mm}\right]  \rule{0mm}{5mm}\right\} \ . \nonumber \\
\end{eqnarray} 
\end{scriptsize}
From these terms one can readily obtain the $h_ih_jh_k$ interactions in the mass eigenstate basis $(h_1, h_2, h_3)$ using the orthogonal matrix $R$ from Eq.~(\ref{R}). In particular, the $g_{i11} (i=2,3)$ coefficients used in Eq.~(\ref{trihiggs}) are
\begin{eqnarray}
g_{i11} = \frac{1}{2} \sum_{a\leq b\leq c} \frac{\partial^3 \mathcal{L}_{3s}}{\partial H_a \partial H_b \partial H_c} \frac{\partial H_a}{\partial h_1} \frac{\partial H_b}{\partial h_1} \frac{\partial H_c}{\partial h_i} = \frac{1}{2} \sum_{a\leq b\leq c} \frac{\partial^3 \mathcal{L}_{3s}}{\partial H_a \partial H_b \partial H_c} R_{1a} R_{1b} R_{ic}  \ ,
\end{eqnarray} 
where $\{H_a\} = (H_1^0, H_2^0, A^0)$.

\section{Formula for $g-2$}
The magnetic and electric dipole moments of a fermion $f$ correspond to the real and imaginary parts of the Wilson coefficient $c$ of the effective operator
\begin{eqnarray}
\mathcal{L}_{eff} = c \bar f_L \sigma_{\mu\nu} f_R F^{\mu\nu} +{\rm h.c.} \ ,
\end{eqnarray}
where in the Type-I and Type-II 2HDMs we consider the main contributions to the coefficient $c$ that arise from the two-loop Barr-Zee type diagrams.
It is  straightforward to translate the electron EDM results to the corresponding muon anomalous dipole moment.
The prescription for the translation is,
\begin{eqnarray}
a_\mu &=& \frac{2m_\mu^2}{e Q_\mu m_e} \times \left\{\begin{array}{ll}
 d_e^\gamma\left( \begin{array}{c}
c_e \to \tilde c_\mu \\
\tilde c_e \to - c_\mu
\end{array} \right), & \hspace{0.5cm} h\gamma\gamma, hZ\gamma\ {\rm diagrams} \\
d_e^\gamma \left( 
{\rm Im} \left(a_{W^+ H^- h_i} \right) \to  -{\rm Re} \left(a_{W^+ H^- h_i} \right)
\rule{0mm}{5mm}\right), & \hspace{0.5cm} W^\pm H_\mp \gamma\ {\rm diagrams}\,({\rm S}) \\
d_e^\gamma \left( 
{\rm Im} \left(c_{\bar t_R b_L H^+}^* c_{\bar \nu e_R H^+}\right) \rule{0mm}{5mm}\right. \nonumber \\
\hspace{2.1cm}\left.\to  -{\rm Re} \left(c_{\bar t_R b_L H^+}^* c_{\bar \nu e_R H^+}\right)
\rule{0mm}{5mm}\right), & \hspace{0.5cm} W^\pm H_\mp \gamma\ {\rm diagrams}\,({\rm F}) \\
\end{array}\right. \nonumber \\
\end{eqnarray}
where $AB\gamma$ corresponds to those Barr-Zee diagrams with $h_1$ lines connected to the upper loop, and the S/F in the bracket 
corresponds to heavy Higgs scalars/SM fermions running in the upper loop.
The $h\gamma\gamma, hZ\gamma$ and $W^\pm H_\mp \gamma\ {\rm diagram}\,({\rm S})$ contributions to the
 EDM have been summarized in Refs.~\cite{Inoue:2014nva, Abe:2013qla}.
The $W^\pm H_\mp \gamma\ {\rm diagram}\,({\rm F})$ contributions to the EDM vanish in 2HDMs with approximate $Z_2$ symmetry, but 
have been calculated in a more general framework in Ref. \cite{BowserChao:1997bb}.  We perform the above translation based on results in Ref. \cite{BowserChao:1997bb}.
See also Ref. \cite{Ilisie:2015tra} for a recent work on $g-2$ in a  2HDM.

We list  below the analytic results for the contributions to the muon $g-2$ in a  2HDM:
\begin{footnotesize}
\begin{eqnarray}
\left(\Delta a_\mu \right)^{h\gamma\gamma}_{f} &=& \frac{G_F m_\mu^2 N_cQ_{f}^2 \alpha}{2\sqrt{2}\pi^3} \sum_{i=1}^3 \left[ - c_{f,i} c_{\mu,i} f(z^i_f) + \tilde c_{f,i} \tilde c_{\mu,i} g(z^i_f) \right] \ , \nonumber \\
\left(\Delta a_\mu \right)^{hZ\gamma}_{f} &=& \frac{G_F m_\mu^2 N_c Q_{t} g_{Z\bar ee}^V g_{Z\bar f f}^V}{8\sqrt{2}\pi^4 Q_\mu} \sum_{i=1}^3 \left[ - c_{f,i} c_{\mu,i} \tilde f \left(z^i_f, \frac{m_f^2}{M_Z^2} \right) + \tilde c_{f,i} \tilde c_{\mu,i} \tilde g \left(z^i_f, \frac{m_f^2}{M_Z^2} \right) \right] \ , \nonumber \\
\left(\Delta a_\mu \right)^{h\gamma\gamma}_W &=& \frac{G_F m_\mu^2 \alpha}{8\sqrt{2}\pi^3} \sum_{i=1}^3 \left[ \left( 6 + \frac{1}{z^i_w} \right) f(z^i_w) + \left( 10 - \frac{1}{z^i_w} \right) g(z^i_w) \right]  (- c_{\mu,i})a_i \ , \nonumber \\
\left(\Delta a_\mu \right)^{hZ\gamma}_W &=& \frac{g_{Z\bar f f}^V g_{ZWW}}{Q_\mu} \frac{G_F m_\mu^2}{32\sqrt{2}\pi^4} \sum_{i=1}^3 \left[ \left(6 -\sec^2\theta_W + \frac{2-\sec^2\theta_W}{2z^i_w} \right)\tilde f(z^i_w, \cos^2\theta_W) \right. \nonumber \\
&&\hspace{1.8cm}+ \left. \left( 10- 3\sec^2\theta_W - \frac{2-\sec^2\theta_W}{2z^i_w}\right)\tilde g(z^i_w, \cos^2\theta_W) \right] (- c_{\mu,i}) a_i \ , \nonumber\\
\left(\Delta a_\mu \right)^{h\gamma\gamma}_{H^+} &=& \frac{G_F m_\mu^2 \alpha}{8\sqrt{2}\pi^3} \left(\frac{v}{m_{H^+}}\right)^2 \sum_{i=1}^3\left[ f(z^i_H) - g(z^i_H)\right]  (- c_{\mu,i}) \bar \lambda_i \nonumber \\
\left(\Delta a_\mu \right)^{hZ\gamma}_{H^+} &=& \frac{g_{Z\bar f f}^V g_{ZH^+ H^-}}{Q_\mu} \frac{G_F m_\mu^2}{32\sqrt{2}\pi^4} \left( \frac{v}{m_{H^+}}\right)^2 \sum_{i=1}^3 \left[ \tilde f(z^i_H, m_{H^+}^2/M_Z^2) - \tilde g(z^i_H, m_{H^+}^2/M_Z^2)\right] (- c_{\mu,i})  \bar \lambda_i \ , \nonumber\\
\left( \Delta a_\mu \right)^{HW\gamma}_H &=& -\frac{G_F m_\mu^2 c_{H^+\bar \nu e^-}}{64\sqrt{2}\pi^4 Q_\mu} \sum_i \left[ \frac{e^2}{2\sin^2\theta_W} \mathcal{I}_4(m_{h_i}^2, m_{H^+}^2) a_i - \mathcal{I}_5(m_{h_i}^2, m_{H^+}^2) \bar \lambda_i \right] (-{\rm Re} \left(a_{W^+ H^- h_i} \right)) \ , \nonumber \\
\left( \Delta a_\mu \right)^{HW\gamma}_{t,b} &=& \left( \frac{3 g^2}{16 \pi^2} \right) \left( \frac{g^2 m_\mu^2}{32 \pi^2 M_W^2} \right) \left(- {\rm Re} \left(c_{\bar t_R b_L H^+}^* c_{\bar H^+\nu e^-}\right) \right) \left(\frac{2}{3} F_t - \frac{1}{3} F_b\right) \ ,
\end{eqnarray}
\end{footnotesize}
where $z^i_{f}= m_{f}^2/m_{h_i}^2$ $(f=t,b)$, $z^i_w = M_W^2/m_{h_i}^2$, $z^i_H = m_{H^+}^2/m_{h_i}^2$, and $c_{e, i}= c_{\mu, i} = c_{\tau, i}$,
$\tilde c_{e, i}= \tilde c_{\mu, i} = \tilde c_{\tau, i}$ can be obtained from Table~\ref{Hcouplings}.

The relevant coefficients are, 
\begin{eqnarray}
g_{Zf\bar f}^V &=& \frac{g}{2\cos\theta_W} \left(T_3^f - 2 Q^f \sin^2\theta_W\right) \ , \nonumber \\
g_{WWZ} &=& e \cot\theta_W \ , \nonumber \\
g_{ZH^+H^-} &=& \frac{1}{2} e \cot\theta_W(1-\tan^2\theta_W) \ , \nonumber \\
\bar\lambda_i 
&=& R_{i1} \cdot \left( \lambda_3 \cos^2\beta + (\lambda_1 - \lambda_4 - {\rm Re}\lambda_5) \sin^2\beta \right) \cos\beta \nonumber \\ 
&& + R_{i2} \cdot \left( \lambda_3 \sin^2\beta + (\lambda_2 - \lambda_4 - {\rm Re}\lambda_5) \cos^2\beta \right) \sin \beta \nonumber \\ 
&& + R_{i3} \cdot {\rm Im} \lambda_5 \, \sin\beta \cos\beta \ , \nonumber \\
a_{W^+ H^- h_i} &=& -\sin\beta R_{i1} + \cos\beta R_{i2} + i R_{i3} \ , \nonumber \\
c_{\bar t_R b_L H^+} &=& \cot\beta \ , \nonumber \\
c_{H^+\bar \nu e^-} &=& \left\{ \begin{array}{ll}
 \cot\beta & \hspace{0.5cm} {\rm Type\ I}  \\
 -\tan\beta & \hspace{0.5cm} {\rm Type\ II}
\end{array}
\right.
\end{eqnarray}

The relevant loop functions are,
\begin{eqnarray}
%I(z) &=& z \int_{0}^1 dx \frac{x^2}{zx^2-x+1} \ , \\
h_0(z) &=& \frac{z^4}{2} \int_0^1 dx \int_0^1 dy \frac{x^3 y^3 (1-x)}{(z^2x(1-xy)+(1-y)(1-x))^2} \ ,\nonumber \\
f(z) &=& \frac{z}{2} \int_0^1 dx \frac{1-2x(1-x)}{x(1-x)-z} \log\frac{x(1-x)}{z} \ ,\nonumber\\ 
g(z) &=& \frac{z}{2} \int_0^1 dx \frac{1}{x(1-x)-z} \log\frac{x(1-x)}{z} \ ,\nonumber\\ 
h(z) &=& \frac{z}{2} \int_0^1 dx \frac{1}{z-x(1-x)} \left( 1+ \frac{z}{z-x(1-x)} \log\frac{x(1-x)}{z} \right) \ , \nonumber\\
\tilde f(x,y) &=& \frac{yf(x)}{y-x} + \frac{x f(y)}{x-y}  \ , \nonumber\\
\tilde g(x,y) &=& \frac{yg(x)}{y-x} + \frac{x g(y)}{x-y} \ , \nonumber\\
\mathcal{I}_{4,5}(m_1^2, m_2^2) &=& \frac{M_W^2}{m_{H^+}^2 - M_W^2} \left(I_{4,5}(M_W^2, m_1^2)-I_{4,5}(m_2^2, m_1^2)\right) \ , \nonumber\\
I_4(m_1^2,m_2^2) &=& \int_0^1 dz (1-z)^2 \left( z-4 + z\frac{m_{H^+}^2 -m_2^2}{M_W^2} \right)  \nonumber \\
&&\hspace{0.6cm} \times \frac{m_1^2}{M_W^2(1-z)+m_2^2 z - m_1^2z(1-z)} \log\frac{M_W^2(1-z)+m_2^2z}{m_1^2z(1-z)} \ ,\nonumber\\ 
I_5(m_1^2,m_2^2) &=& \int_0^1 dz \frac{m_1^2 z(1-z)^2}{M_{H^+}^2(1-z)+m_2^2 z - m_1^2z(1-z)} \log\frac{M_{H^+}^2(1-z)+m_2^2z}{m_1^2z(1-z)} \ ,\nonumber\\
Sp(z) &=& - \int_0^z t^{-1} \ln(1-t) dt \ , \nonumber\\
T(z) &=& \frac{1-3z}{z^2} \frac{\pi^2}{6} - \left( \frac{1}{z} - \frac{5}{2} \right) \ln z - \frac{1}{z} - \left( 2 - \frac{1}{z} \right)\left( 1 - \frac{1}{z} \right) Sp(1-z) \ , \nonumber \\
B(z) &=& \frac{1}{z} + \frac{2z-1}{z^2} \frac{\pi^2}{6}  + \left( \frac{3}{2} - \frac{1}{z} \right) \ln z - \left( 2 - \frac{1}{z} \right) \frac{1}{z} Sp(1-z) \ , \nonumber \\
F_t &=& \frac{T(m_{H^+}^2/m_t^2) - T(M_W^2/m_t^2)}{m_{H^+}^2/m_t^2 - M_W^2/m_t^2} \ , \nonumber \\
F_b &=& \frac{B(m_{H^+}^2/m_t^2) - B(M_W^2/m_t^2)}{m_{H^+}^2/m_t^2 - M_W^2/m_t^2} \ .
\end{eqnarray}

\section*{Acknowledgements}
We thank Jing Shu and  Michael Spira for useful discussions.
The work of C.-Y. Chen and S. Dawson is supported by the U.S. Department of Energy under grant No. DE-AC02-98CH10886 and contract DE-AC02-76SF00515. This work of Y. Zhang is supported by the Gordon and Betty Moore Foundation through Grant \#776 to the Caltech Moore Center for Theoretical Cosmology and Physics, and by the DOE Grant DE-FG02-92ER40701, and also by a DOE Early Career Award under Grant No. DE-SC0010255.

\end{document}